\documentclass[%
 reprint,
 twocolumn,
 showpacs,
 aps,
]{revtex4-1}

\usepackage{
amsmath, 
amssymb,
empheq,
graphicx, 
dcolumn, 
bm, 
verbatim, 
color, 
subfigure, 
hyperref, 
multirow
}

\usepackage{ulem}
\definecolor{rgbred}{rgb}{1,0,0}
\definecolor{rgbgreen}{rgb}{0,0.5,0}
\definecolor{rgbblue}{rgb}{0,0,1}

\allowdisplaybreaks[4]

\makeatletter
\renewcommand{\dddot}[1]{%
{\mathop{#1\hspace{0pt}}\limits^{\vbox to-1.4\ex@{\kern-\tw@\ex@
\hbox {\normalfont .\kern-.1em.\kern-.1em.}\vss}}}}
\makeatother

\newcommand{\1}{\mbox{1}\hspace{-0.25em}\mbox{l}}
\newcommand{\ket}[1]{| #1 \rangle}
\newcommand{\bra}[1]{\langle #1 |}
\newcommand{\submin}[1]{\langle #1 \rangle}
\newcommand{\bket}[2]{\langle #1 | #2 \rangle}
\newcommand{\bketm}[3]{\langle #1 | #2 | #3 \rangle}
\newcommand{\diff}[2]{\frac{{\rm d}#1}{{\rm d}#2}}
\newcommand{\pdiff}[2]{\frac{\partial #1}{\partial #2}}

\begin{document}
\preprint{APS/123-QED}

\title{Transition probability generating function of a transitionless quantum parametric oscillator}

\author{Hiroaki Mishima}
\thanks{mishima.hiroaki@j.mbox.nagoya-u.ac.jp}
\author{Yuki Izumida}
\thanks{Present address: Department of Complex Systems Science, Graduate School of Informatics, Nagoya University, Nagoya 464-8601, Japan; Electronic address: izumida@i.nagoya-u.ac.jp}%
\affiliation{%
Department of Complex Systems Science, Graduate School of Information Science, Nagoya University, Nagoya 464-8601, Japan
}%
\date{\today}

\begin{abstract}
The transitionless tracking (TT) algorithm enables the exact tracking of quantum adiabatic dynamics in an arbitrary short time by adding a counterdiabatic Hamiltonian to the original adiabatic Hamiltonian.
By applying Husimi's method originally developed for a quantum parametric oscillator (QPO) to the transitionless QPO achieved using the TT algorithm, 
we obtain the transition probability generating function with a time-dependent parameter constituted with solutions of the corresponding classical parametric oscillator (CPO).
By obtaining the explicit solutions of this CPO using the phase-amplitude method, we find that the time-dependent parameter can be reduced to the frequency ratio between the Hamiltonians without and with the counterdiabatic Hamiltonian, from which we can easily characterize the result achieved by the TT algorithm.
We illustrate our theory by showing the trajectories of the CPO on the classical phase space, which elucidate the effect of the counterdiabatic Hamiltonian of the QPO.
\end{abstract}


\maketitle

\section{Introduction}

Suppose that we can change a parameter of a system to control it.
The dynamics of the system under a change of the control parameter that is slow enough compared to the intrinsic time scale of the system is called an adiabatic process . 
An adiabatic invariant is a quantity that is conserved in the limit of infinitely slow change of the control parameter.
Adiabatic invariants appear in both classical and quantum mechanics.
A classical example of an adiabatic invariant is the area enclosed by a trajectory in a classical phase space.
A quantum analog of the adiabatic invariant is the principal quantum number, which labels different energy levels.
Ideally, a quantum system exhibits no transition between energy levels during an adiabatic process.
However, in a realistic process carried out for a finite duration, the adiabatic invariant is not conserved
and transition between different energy levels occurs in a quantum scenario.

A controlled quantum system is described as follows.
Suppose that the system obeys a Hamiltonian $\hat H^{\rm ad}_t=\hat H^{\rm ad}(\lambda_t)$, which is a function of an external time-dependent parameter $\lambda_t$.
The instantaneous eigenstate $\ket{n;\lambda_t}$ satisfies $\hat H^{\rm ad}_t\ket{n;\lambda_t}=E_{n,t}\ket{n;\lambda_t}$.
The quantum adiabatic theorem~\cite{Born, Kato} ensures that the solution of the time-dependent Schr\"odinger equation is approximated with the instantaneous eigenstate if the initial state is an instantaneous eigenstate and the parameter $\lambda_t$ varies slowly enough.
Under this adiabatic approximation, the solution of the Schr\"odinger equation ${\rm i}\hbar\diff{}{t}\ket{\psi_n(t)}=\hat{H}^{\rm ad}_t\ket{\psi_n(t)}$ is given as
\begin{align}
\ket{\psi_n(t)}\simeq e^{{\rm i}\xi_{n,t}}\ket{n;\lambda_t},
\label{eq:psi_qadt}
\end{align}
where the phase $\xi_{n,t} \in\mathbb R$ is
\begin{align}
\xi_{n,t}
\equiv
-\frac{1}{\hbar}\int_{t_0}^t{\rm d}t'E^{\rm ad}_{n,t'}+{\rm i}\int_{t_0}^t{\rm d}t'\bketm{n;\lambda_{t'}}{\diff{}{t'}}{n;\lambda_{t'}}.
\label{eq:dyn_berry_ph}
\end{align}
The first and second terms of $\xi_{n,t}$ are the dynamical phase and geometric phase, respectively.

Husimi described this quantum adiabatic theorem in terms of {\it classical} adiabatic invariants~\cite{Husimi}.
For a quantum parametric oscillator (QPO), he developed a method with which the propagator of the quantum system can be expressed using linearly independent solutions of equations of motion for a corresponding classical parametric oscillator (CPO).
From the propagator, one can calculate transition probabilities between two arbitrary states from a transition probability generating function . Husimi found that the transition probability generating function of a QPO is characterized by a parameter $Q^*_t$. 
Moreover, he found that the value of $Q^*_t$ is unity if and only if no transitions occur between arbitrary instantaneous eigenstates. Hence, we call $Q^*_t$ as Husimi's measure of adiabaticity.
Husimi also found that $Q^*_t$ is a function of two adiabatic invariants of a CPO [Eq.~(\ref{eq:husimi})].
Each adiabatic invariant is defined in terms of each solution of a CPO [Eqs.~(\ref{eq:cpo_mu_omega}) and (\ref{eq:cpo_nu_omega})].

Naively, an adiabatic process takes an infinitely long time. Hence, it is natural to seek a method for achieving the same final state of this process in a finite duration. Various methods have been proposed as shortcuts to adiabaticity~\cite{Torrontegui13}, such as the assisted adiabatic passage~\cite{Demirplak03,Demirplak05}, transitionless tracking algorithm~\cite{Berry09}, fast-forward method~\cite{Masuda&Nakamura08,Masuda&Nakamura10,Torrontegui12,Masuda14,Jarzynski17}, Lewis-Riesenfeld invariant-based inverse engineering~\cite{LR69,Chen10}, scale-invariant driving~\cite{Deffner14}, generator of adiabatic transport~\cite{Jarzynski13}, and quantum brachistochrone~\cite{Carlini06,Carlini07,Takahashi13}.
These methods have received much attention recently for both theoretical interest and experimental relevance.

Among the various methods, the transitionless tracking (TT) algorithm introduces a counterdiabatic Hamiltonian $\hat H^{\rm cd}_t$ 
for canceling the deviation from exact tracking along instantaneous eigenstates of the original adiabatic Hamiltonian $\hat H^{\rm ad}_t$~\cite{Berry09}.
The counterdiabatic Hamiltonian is defined as
\begin{align}
\hat{H}^{\rm cd}_t\equiv {\rm i}\hbar\sum_{n=0}^\infty(\hat{\1}-\ket{n;\lambda_t}\bra{n;\lambda_t})\diff{\ket{n;\lambda_t}}{t}\bra{n;\lambda_t}.
\end{align}
In this method, it is assumed that the system obeys the total Hamiltonian $\hat H ^{\rm TT}_t\equiv\hat H^{\rm ad}_t+\hat H^{\rm cd}_t$,
which we call the TT Hamiltonian.
Then, the state vector of Eq.~(\ref{eq:psi_qadt}) is an exact solution of the Schr\"odinger equation for the Hamiltonian $\hat H^{\rm TT}_t$.
For a case in which the original Hamiltonian $\hat H^{\rm ad}_t$ is a QPO, the counterdiabatic Hamiltonian $\hat H^{\rm cd}_t$ has been calculated explicitly~\cite{Muga}. 
Hence, the exact tracking of adiabatic dynamics in an arbitrary short time has been achieved for a QPO.

On the other hand, 
Husimi showed that an adiabatic process of a usual QPO is characterized by the transition probability generating function
with a parameter that is a function of adiabatic invariants.
Then, it is natural to ask what type of parameter characterizes the 
adiabatic process of the transitionless QPO
driven by the TT Hamiltonian including the counterdiabatic term.
For answering this question, it is necessary to calculate the probability generating function of the transitionless QPO 
by applying Husimi's method to this system.

In the present study, we characterize the transitionless QPO with the TT Hamiltonian 
by using a transition probability generating function with a new parameter.
By introducing an instantaneous eigenstate of the TT Hamiltonian,
we apply Husimi's method to the transitionless QPO to obtain the propagator expressed with independent solutions of the corresponding CPO. 
By using this propagator, we obtain the probability generating function with the time-dependent parameter as the main result [Eqs.~(\ref{eq:pgf_TT}) and (\ref{eq:Q_TT})], from which the adiabatic process in an arbitrary short time achieved by the TT algorithm is easily characterized.
We obtain this parameter by solving the equations of the CPO by using the phase-amplitude method~\cite{Andersson}.
We illustrate our theory by exhibiting some trajectories of the solutions of the CPO of a specific case, 
which visualize the effect of the counterdiabatic term of the QPO on the classical phase space.

The remainder of this paper is organized as follows.
In Sec.~\ref{sec:pre}, we introduce the transitionless QPO and its propagator based on Husimi's method. In Sec.~\ref{sec:results}, we present the 
transition probability generating function with the new parameter 
that characterizes the TT algorithm as our main result. A specific case is also presented in this section to illustrate our theory.
We conclude the paper in Sec.~\ref{sec:sum}.

\section{Preliminaries}
\label{sec:pre}

\subsection{Transitionless quantum parametric oscillator}
\label{sec:ttqpo}

Let $\omega_t, M, \hat{x},$ and $\hat{p}$ be, respectively, the frequency at time $t$, mass, position operator, and momentum operator, where $\hat x$ and $\hat p$ satisfy the canonical commutation relation $[\hat x,\hat p]={\rm i}\hbar$.
For the QPO, the TT Hamiltonian $\hat{H}^\text{TT}_t$ is given by~\cite{Muga}
\begin{align}
\hat{H}^\text{TT}_t
&=
\frac{\hat{p}^2}{2M}+\frac{M}{2}\omega^2_t\hat{x}^2\vphantom{\frac{\dot\omega_t}{\omega_t}}-\frac{1}{2}\frac{\dot\omega_t}{\omega_t}\frac{\hat{x}\hat{p}+\hat{p}\hat{x}}{2},
\label{eq:hamiltonian_tt}
\end{align}
where the first two terms are part of the adiabatic Hamiltonian $\hat{H}^\mathrm{ad}_t$ 
and the third term is part of the counterdiabatic Hamiltonian $\hat{H}^\mathrm{cd}_t$.
Hereafter, we denote the time derivative by a dot.

We rewrite $\hat{H}^\text{TT}_t$ of the QPO in Eq.~(\ref{eq:hamiltonian_tt}) with the instantaneous ladder operator $\hat{b}_t$ as
\begin{align}
\hat{H}^\text{TT}_t
&=
\hbar\varOmega_t\biggl(\hat{b}^\dagger_t\hat{b}_t+\frac{1}{2}\biggr),
\label{eq:hamiltonian _tt_r}
\end{align}
where 
\begin{align}
&\varOmega_t
\equiv 
\sqrt{\omega^2_t-\frac{1}{4}\frac{\dot\omega_t^2}{\omega_t^2}},
\label{eq:Omega}
\\
&\hat{b}_t
\equiv 
\sqrt{\frac{M\varOmega_t}{2\hbar}}\biggl(\zeta_t\hat{x}+\frac{{\rm i}\hat{p}}{M\varOmega_t}\biggr),
\label{eq:boson_op}
\end{align}
with $\zeta_t\equiv 1+\frac{1}{2{\rm i}\varOmega_t}\frac{\dot\omega_t}{\omega_t}$.
Since $\hat{b}_t$ satisfies the Boson commutation relation $[\hat{b}_t,\hat{b}^\dagger_t]=1$, $\hat{H}^\text{TT}_t$ can be 
regarded as the Hamiltonian of a certain type of harmonic oscillator with the energy-level interval $\hbar\varOmega_t$ under the assumption of $\varOmega_t >0$.
We adopt the Schr\"odinger picture to interpret these ladder operators $\hat{b}_t$ and $\hat{b}^\dagger_t$ at time $t$.
They annihilate and create different instantaneous eigenstates, respectively, and do not commute in general with different time labels.

Let $\ket{n;\varOmega_t}$ be an instantaneous $n$-th excited energy eigenstate 
of $\hat{H}_t^{\rm TT}$ in Eq.~(\ref{eq:hamiltonian _tt_r}) that satisfies
$\sqrt{n!}\ket{n;\varOmega_t}=\hat{b}_t^{\dagger n}\ket{0;\varOmega_t}$
and
$\hat{b}^\dagger_t\hat{b}_t\ket{n;\varOmega_t}=n\ket{n;\varOmega_t}$,
where the vacuum state $\ket{0;\varOmega_t}$ is defined as $\hat{b}_t\ket{0;\varOmega_t}\equiv 0$.
The instantaneous $n$-th excited energy eigenfunction for position $x$ is given by (see Appendix~\ref{sec:deriv_energy_ef})
\begin{align}
&\bket{x}{n;\varOmega_t}
\notag\\
&=
\frac{1}{\sqrt{2^n n!}}\biggl(\frac{M\varOmega_t}{\pi\hbar}\biggr)^{1/4}{\rm H}_n\biggl(\sqrt{\frac{M\varOmega_t}{\hbar}}x\biggr)\exp\biggl(-\frac{\zeta_t M\varOmega_t}{2\hbar}x^2\biggr),
\label{eq:efunc_tt}
\end{align}
where ${\rm H}_n(\cdot)$ is the $n$-th-degree Hermite polynomial.

\subsection{Propagator based on Husimi's method}
\label{sec:probability_husimi}

We consider the transition from an initial state $\ket{n;\varOmega_{t_0}}$ 
at initial time $t_0$ to a certain state $\ket{m;\varOmega_t}$ at time $t$ ($t_0 \le t \le t_{\rm f}$),
characterized by the transition probability $P^{m,n}_{t,t_0}$ defined as 
\begin{align}
P^{m,n}_{t,t_0}
\equiv
\biggl|\iint_{\mathbb{R}^2}{\rm d}x{\rm d}x_0\bket{m;\varOmega_t}{x}U_{t,t_0}(x|x_0)\bket{x_0}{n;\varOmega_{t_0}}\biggr|^2,
\label{eq:trans_prob}
\end{align}
where $U_{t,t_0}(x|x_0)$ is the propagator.
The TT algorithm usually imposes the boundary condition $\dot{\omega}_{t_0}=\dot{\omega}_{t_{\rm f}}=0$ 
($\hat{H}^\text{cd}_{t_0}=\hat{H}^\text{cd}_{t_{\rm f}}=0$) at the initial and final times $t=t_0$ and $t_{\rm f}$, respectively, such that
the instantaneous eigenstates of the original Hamiltonian and the TT Hamiltonian coincide at these times.
However, we first consider the transition probability of Eq.~(\ref{eq:trans_prob}) as a more general case and impose this boundary condition later.

By applying Husimi's method~\cite{Husimi}, we can concretely obtain the propagator as (see Appendix~\ref{sec:deriv_prop})
\begin{align}
U_{t,t_0}(x|x_0)
&=
\sqrt{\frac{M}{2\pi{\rm i}\hbar\mu_t}}\exp\biggl[\frac{{\rm i}M}{2\hbar}\biggl\{\biggl(\frac{\dot\mu_t}{\mu_t}+\frac{1}{2}\frac{\dot\omega_t}{\omega_t}\biggr)x^2
\notag\\
&\hphantom{=}-\frac{2xx_0}{\mu_t}+\biggl(\frac{\nu_t}{\mu_t}-\frac{1}{2}\frac{\dot\omega_{t_0}}{\omega_{t_0}}\biggr)x_0^2\biggr\}\biggr],
\label{eq:gf_tt}
\end{align}
where $\mu_t$ and $\nu_t$ are the solutions of the CPO with different initial conditions at $t=t_0$. They are given by 
\begin{align}
\ddot\mu_t+\tilde\varOmega^2_t\mu_t
&=
0, &\mu_{t_0}=0,&&\dot\mu_{t_0}=1,
\label{eq:cpo_mu}
\\
\ddot\nu_t+\tilde\varOmega^2_t\nu_t
&=
0, &\nu_{t_0}=1,&&\dot\nu_{t_0}=0,
\label{eq:cpo_nu}
\end{align}
where
\begin{align}
\tilde\varOmega_t
\equiv
\sqrt{\varOmega^2_t+\frac{1}{2}\diff{}{t}\frac{\dot\omega_t}{\omega_t}}
=
\sqrt{\omega^2_t-\frac{3}{4}\frac{\dot\omega^2_t}{\omega^2_t}+\frac{1}{2}\frac{\ddot\omega_t}{\omega_t}}.
\label{eq:omega_tilde}
\end{align}
We can confirm that the solutions $\mu_t$ and $\nu_t$ satisfying Eqs.~(\ref{eq:cpo_mu}) and (\ref{eq:cpo_nu}) are linearly independent because the Wronskian $W_t$ is a non-zero constant for arbitrary time $t$:
\begin{align}
W_t
\equiv
\dot\mu_t\nu_t-\mu_t\dot\nu_t
=
1.
\label{eq:wronskian}
\end{align}
We note that $\mu_t$ has the dimension of time, whereas $\nu_t$ is dimensionless.

\section{Main results}
\label{sec:results}

\subsection{Transition probability generating function}
\label{sec:pgf}

While the concrete expression for the transition probabilities in Eq.~(\ref{eq:trans_prob}) is complicated, a calculation of the generating function,
\begin{align}
\mathcal{P}^{u,v}_{t,t_0}\equiv \sum_{n,m=0}^\infty u^n v^m P^{m,n}_{t,t_0},
\label{eq:def_pgf}
\end{align}
yields a rather simple expression (see Appendix~\ref{sec:deriv_pgf}):
\begin{align}
\mathcal{P}^{u,v}_{t,t_0}
=
\sqrt{\frac{2}{Q^{\rm TT}_t(1-u^2)(1-v^2)+(1+u^2)(1+v^2)-4uv}},
\label{eq:pgf_TT}
\end{align}
where $Q^{\rm TT}_t$ is a time-dependent parameter defined as
\begin{align}
Q^{\rm TT}_t
&\equiv 
\varOmega_{t_0}\frac{\frac{1}{2}(\dot\mu^2_t+\varOmega^2_t\mu^2_t)}{\varOmega_t}+\varOmega_{t_0}^{-1}\frac{\frac{1}{2}(\dot\nu^2_t+\varOmega^2_t\nu^2_t)}{\varOmega_t}
\notag\\
&\hphantom{=}
+\frac{\dot\omega_t}{\omega_t}\frac{\varOmega_{t_0}^2\dot\mu_t\mu_t+\dot\nu_t\nu_t+\frac{1}{2}\frac{\dot\omega_t}{\omega_t}(\varOmega_{t_0}^2\mu_t^2+\nu_t^2)}{\varOmega_t\varOmega_{t_0}}.
\label{eq:Q_TT_general}
\end{align}
We can simplify this parameter 
by imposing $\dot\omega_{t_0}=0$ ($\hat{H}^\text{cd}_{t_0}=0$), that is, by preparing $\ket{n;\varOmega_{t_0}}=\ket{n;\omega_{t_0}}$ as the initial state.
In this case, we can obtain the solutions of $\mu_t$ and $\nu_t$ in Eqs.~(\ref{eq:cpo_mu}) and (\ref{eq:cpo_nu}) as 
\begin{align}
\mu_t&=\frac{1}{\sqrt{\omega_{t_0}\omega_t}}\sin\biggl(\int_{t_0}^t{\rm d}t'\omega_{t'}\biggr)&
&;\,\dot\omega_{t_0}=0,
\label{eq:mu_simple_sol}\\
\nu_t&=\sqrt{\frac{\omega_{t_0}}{\omega_t}}\cos \biggl(\int_{t_0}^t{\rm d}t'\omega_{t'}\biggr)&
&;\,\dot\omega_{t_0}=0,
\label{eq:nu_simple_sol}
\end{align}
respectively; these solutions will be derived in Sec.~\ref{derivation}.
Then, by substituting Eqs.~(\ref{eq:mu_simple_sol}) and (\ref{eq:nu_simple_sol}) into Eq.~(\ref{eq:Q_TT_general}), 
we find that the third term in Eq.~(\ref{eq:Q_TT_general}) vanishes and that the former two terms become the ratio of the frequency of the adiabatic Hamiltonian to that of the TT Hamiltonian defined in Eq.~(\ref{eq:Omega}):
\begin{align}
Q^{\rm TT}_t
&=
\frac{\omega_t}{\varOmega_t}
\quad ;\,\dot\omega_{t_0}=0.
\label{eq:Q_TT}
\end{align}
The key to obtain this simple form is the explicit solutions Eqs.~(\ref{eq:mu_simple_sol}) and (\ref{eq:nu_simple_sol}), despite the time dependence of $\omega_t$.
The probability generating function defined by Eq.~(\ref{eq:pgf_TT}) with the simple time-dependent parameter expressed by
Eq.~(\ref{eq:Q_TT}) is the main result of the present paper.

By taking the first derivative of Eq.~(\ref{eq:pgf_TT}) with respect to $v$ and substituting $v=1$, we obtain
\begin{align}
\sum_{n=0}^\infty u^n\sum_{m=0}^\infty m P_{t,t_0}^{m,n}
=
\pdiff{\mathcal{P}^{u,v}_{t,t_0}}{v}\biggr|_{v=1}
=
\frac{Q^{\rm TT}_t(1+u)-(1-u)}{2(1-u)^2}.
\label{eq:pgf_TT_u}
\end{align}
By expanding the right-hand side of Eq.~(\ref{eq:pgf_TT_u}) with respect to $u$, we can show the following relation between $Q^{\rm TT}_t$ and the mean quantum number 
$\submin{m}_{n,t}\equiv\sum_{m=0}^\infty mP^{m,n}_{t,t_0}$~\cite{Husimi, Deffner08}:
\begin{align}
Q^{\rm TT}_t
=
\frac{\submin{m}_{n,t}+\frac{1}{2}}{n+\frac{1}{2}}.
\label{eq:Q-mean_m}
\end{align}
When $Q^{\rm TT}_{t}=1$, we find $\submin{m}_{n,t}=n$ and can show $\mathcal{P}^{u,v}_{t,t_0}|_{Q^{\rm TT}_t=1}=\frac{1}{1-uv}$, 
from which $P^{m,n}_{t,t_0}=\delta_{mn}$ (no transition) follows.
Indeed, by letting $\ket{n;\varOmega_{t_{\rm f}}}=\ket{n;\omega_{t_{\rm f}}}$ 
be the final state by imposing $\dot{\omega}_{t_{\rm f}}=0$ ($\hat{H}^\text{cd}_{t_{\rm f}}=0$), we find that our new parameter $Q^{\rm TT}_t$ in Eq.~(\ref{eq:Q_TT}) at $t=t_{\rm f}$
is always unity.
This implies $P^{m,n}_{t_{\rm f},t_0}=\delta_{mn}$, and 
the final state with the same quantum number as the initial state is achieved.

\subsection{Comparison with Husimi's measure of adiabaticity}

For a usual QPO in the absence of $\hat{H}^{\rm cd}_t$, the probability generating function 
from an initial state $\ket{n;\omega_{t_0}}$ to a certain state $\ket{m;\omega_t}$ is given by Eq.~(\ref{eq:pgf_TT}) with $Q_t^{\rm TT}$ replaced by Husimi's measure of adiabaticity, $Q^*_t$, defined as~\cite{Husimi} 
\begin{eqnarray}
Q_t^* 
\equiv \omega_{t_0}\frac{E^{(\mu)}_t}{\omega_t}+\omega_{t_0}^{-1}\frac{E^{(\nu)}_t}{\omega_t},
\label{eq:husimi}
\end{eqnarray}
where
\begin{eqnarray}
&&E^{(\mu)}_t\equiv \frac{1}{2}(\dot\mu^2_t+\omega^2_t\mu^2_t),\label{eq:mu_true_energy}\\
&&E^{(\nu)}_t\equiv \frac{1}{2}(\dot\nu^2_t+\omega^2_t\nu^2_t),\label{eq:nu_true_energy}
\end{eqnarray}
are the classical energies of $\mu_t$ and $\nu_t$, respectively, and $\mu_t$ and $\nu_t$ obey the equations of motion for the usual CPO: 
\begin{align}
\ddot\mu_t+\omega^2_t\mu_t
&=
0, &\mu_{t_0}=0,&&\dot\mu_{t_0}=1,
\label{eq:cpo_mu_omega}
\\
\ddot\nu_t+\omega^2_t\nu_t
&=
0, &\nu_{t_0}=1,&&\dot\nu_{t_0}=0.
\label{eq:cpo_nu_omega}
\end{align}
$Q^*_t$ is constituted with a linear combination of these two energies of the CPO divided by the common frequency, $\frac{E^{(\mu)}_t}{\omega_t}$ and $\frac{E^{(\nu)}_t}{\omega_t}$.
During an adiabatic process with the slowly changing frequency $\dot{\omega}_t\simeq 0$,
these quantities are conserved 
as the {\it adiabatic invariants} as $\frac{1}{2\omega_{t_0}}\equiv J^{(\mu)}$ and $\frac{\omega_{t_0}}{2}\equiv J^{(\nu)}$, respectively.
They are equivalent to the areas of the ellipses enclosed by the trajectories of the CPO on the classical phase space with $E_{t_0}^{(\mu)}=\frac{1}{2}$ and $E_{t_0}^{(\nu)}=\frac{\omega_{t_0}^2}{2}$ determined from the initial conditions in Eqs.~(\ref{eq:cpo_mu_omega}) and (\ref{eq:cpo_nu_omega}).
In contrast to Eqs.~(\ref{eq:cpo_mu}) and (\ref{eq:cpo_nu}), which have the explicit solutions given by Eqs.~(\ref{eq:mu_simple_sol}) and (\ref{eq:nu_simple_sol}), respectively, 
we may not obtain explicit solutions for Eqs.~(\ref{eq:cpo_mu_omega}) and (\ref{eq:cpo_nu_omega}).
Therefore, we may not obtain a simple form as in Eq.~(\ref{eq:Q_TT}) for this usual QPO.
However, $Q^*_{t}\simeq 1$ holds during the adiabatic process 
owing to the existence of these adiabatic invariants. This is an expression of the quantum adiabatic theorem as it implies $P^{m,n}_{t,t_0}=\delta_{mn}$ 
for any $t$.

For a comparison with Husimi's measure of adiabaticity, it is convenient to express $Q^{\rm TT}_t$ in Eq.~(\ref{eq:Q_TT}) as
\begin{align}
Q^{\rm TT}_t
&=\omega_{t_0}\frac{\mathcal E^{(\mu)}_t}{\varOmega_t}+\omega_{t_0}^{-1}\frac{\mathcal E^{(\nu)}_t}{\varOmega_t}
\quad ;\,\dot\omega_{t_0}=0,
\label{eq:Q_TT_E}
\end{align}
where we have introduced the two classical ``energies'' as
\begin{eqnarray}
&&\mathcal{E}^{(\mu)}_t\equiv \frac{1}{2}(\dot\mu^2_t+\varOmega^2_t\mu^2_t),\\
&&\mathcal{E}^{(\nu)}_t\equiv \frac{1}{2}(\dot\nu^2_t+\varOmega^2_t\nu^2_t).
\end{eqnarray}
It should be noted that these ``energies'' are defined with the frequency $\varOmega_t$ 
that appeared in the QPO with the TT Hamiltonian in Eq.~(\ref{eq:Omega}), 
but $\mu_t$ and $\nu_t$ are solutions of the equations of the CPO in 
Eqs.~(\ref{eq:cpo_mu}) and (\ref{eq:cpo_nu}) with the frequency $\tilde{\varOmega}_t$.
We can rewrite $\mathcal{E}^{(\mu)}_t$ and $\mathcal{E}^{(\nu)}_t$ as (see Sec.~\ref{derivation})
\begin{align}
\mathcal{E}^{(\mu)}_t
&=
\frac{\omega_t}{2\omega_{t_0}}-\biggl(\dot\mu_t+\frac{\mu_t}{2}\frac{\dot\omega_t}{\omega_t}\biggr)\frac{\mu_t}{2}\frac{\dot\omega_t}{\omega_t}&
&;\,\dot\omega_{t_0}=0,
\label{eq:energy_mu}\\
\mathcal{E}^{(\nu)}_t
&=
\frac{\omega_{t_0}\omega_t}{2}-\biggl(\dot\nu_t+\frac{\nu_t}{2}\frac{\dot\omega_t}{\omega_t}\biggr)\frac{\nu_t}{2}\frac{\dot\omega_t}{\omega_t}&
&;\,\dot\omega_{t_0}=0,
\label{eq:energy_nu}
\end{align}
respectively.
Because $\mathcal{E}^{(\mu)}_{t_0}=\frac{1}{2}$
and $\mathcal{E}^{(\nu)}_{t_0}=\frac{\omega^2_{t_0}}{2}$ for $\dot\omega_{t_0}=\dot\omega_{t_{\rm f}}=0$ readily follow from the initial conditions in Eqs.~(\ref{eq:cpo_mu}) and (\ref{eq:cpo_nu}),
by using Eqs.~(\ref{eq:energy_mu}) and (\ref{eq:energy_nu}), we can also show that $\mathcal{E}^{(\mu)}_{t_{\rm f}}=\frac{\omega_{t_{\rm f}}}{2\omega_{t_0}}$
and
$\mathcal{E}^{(\nu)}_{t_{\rm f}}=\frac{\omega_{t_0}\omega_{t_{\rm f}}}{2}$.
Therefore, we obtain $\frac{\mathcal{E}^{(\mu)}_{t_0}}{\omega_{t_0}}=\frac{\mathcal{E}^{(\mu)}_{t_{\rm f}}}{\omega_{t_{\rm f}}}=J^{(\mu)}$
and
$\frac{\mathcal{E}^{(\nu)}_{t_0}}{\omega_{t_0}}=\frac{\mathcal{E}^{(\nu)}_{t_{\rm f}}}{\omega_{t_{\rm f}}}=J^{(\nu)}$, i.e., at both the initial and final times, the values of these ``energies" divided by the common frequency $\varOmega_t$ agree with 
the adiabatic invariants $J^{(\mu)}$ and $J^{(\nu)}$. This explains the reason for $Q^{\rm TT}_{t_{\rm f}}=1$ 
at the final time $t_{\rm f}$ in a manner comparable to Husimi's measure of adiabaticity.
During the intermediate times, however, $Q^{\rm TT}_t=1$ does not hold in general, 
because the exact solution Eq.~(\ref{eq:psi_qadt}) for the TT Hamiltonian may be nonadiabatic 
with respect to the instantaneous eigenstate of this Hamiltonian~\cite{Muga}.
This behavior may be similar to that of the system with the fast-forward method~\cite{Jarzynski17} being applied, 
where the system is allowed to deviate from the original adiabatic path and returns to it only at the end of the process. 
We note that in our case with the TT algorithm the state vector itself always tracks the original adiabatic path given by Eq.~(\ref{eq:psi_qadt}).

\subsection{Derivation of key equations}
\label{derivation}

In this subsection, we derive the key equations Eqs.~(\ref{eq:mu_simple_sol}), (\ref{eq:nu_simple_sol}), (\ref{eq:energy_mu}), and (\ref{eq:energy_nu}).

We first derive Eqs.~(\ref{eq:mu_simple_sol}) and (\ref{eq:nu_simple_sol}) as the solutions of Eqs.~(\ref{eq:cpo_mu}) and (\ref{eq:cpo_nu}), respectively, by using the 
phase-amplitude method~\cite{Andersson}.
We define the time-dependent function $\rho_t$ in terms of the two linearly independent solutions $\mu_t$ and $\nu_t$ satisfying Eqs.~(\ref{eq:cpo_mu}) and (\ref{eq:cpo_nu}), respectively, as
\begin{align}
\rho_t
\equiv
\sqrt{\frac{\varOmega_{t_0}^2\mu_t^2+\nu_t^2}{\varOmega_{t_0}}}.
\label{eq:def_rho}
\end{align}
We can then rewrite the Wronskian in Eq.~(\ref{eq:wronskian}) as follows by eliminating either $\mu_t$ or $\nu_t$ from Eq.~(\ref{eq:wronskian}):
\begin{empheq}[left={W_t=\empheqlbrace}]{align}
&-\frac{\rho_t}{\sqrt{\varOmega_{t_0}^{-1}\rho_t^2-\mu_t^2}}(\dot\rho_t\mu_t-\rho_t\dot\mu_t)
\equiv
W_t^{(\mu)},
\label{eq:wronskian_mu}\\
&\frac{\rho_t}{\sqrt{\varOmega_{t_0}\rho_t^2-\nu_t^2}}(\dot\rho_t\nu_t-\rho_t\dot\nu_t)
\equiv
W_t^{(\nu)},
\label{eq:wronskian_nu}
\end{empheq}
where we note that $W_t=W_t^{(\mu)}=W_t^{(\nu)}=1$ holds for an arbitrary time $t$.
The time-evolution equation of $\rho_t$ in Eq.~(\ref{eq:def_rho}) is obtained 
by differentiating the Wronskian of Eqs.~(\ref{eq:wronskian_mu}) and (\ref{eq:wronskian_nu}) with respect to time $t$ and by using Eqs.~(\ref{eq:cpo_mu}), (\ref{eq:cpo_nu}), and (\ref{eq:wronskian}) (see Appendix~\ref{sec:deriv_ermakoveq}):
\begin{align}
\ddot\rho_t+\tilde\varOmega_t^2\rho_t
=
\frac{W_t^2}{\rho_t^3},
\label{eq:Ermakov_eq}
\end{align}
which is called the Ermakov equation~\cite{Ermakov,Pinney50,Leach08}.
On the other hand, by integrating Eqs.~(\ref{eq:wronskian_mu}) and (\ref{eq:wronskian_nu}), we obtain
\begin{align}
\mu_t&=\frac{\rho_t}{\sqrt{\varOmega_{t_0}}}\sin \theta_t,\label{eq:mu_ermakov_formal}\\
\nu_t&=\sqrt{\varOmega_{t_0}}\rho_t\cos \theta_t,\label{eq:nu_ermakov_formal}
\end{align}
where
\begin{eqnarray}
\theta_t
\equiv
\int_{t_0}^t{\rm d}t'\frac{W_{t'}^{(\mu)}}{\rho_{t'}^2}=\int_{t_0}^t{\rm d}t'\frac{W_{t'}^{(\nu)}}{\rho_{t'}^2}
\end{eqnarray}
is a phase function (see Appendix~\ref{sec:deriv_ermakoveq_phamp}).
This description of the coordinate variables $\mu_t$ and $\nu_t$ in terms of $\rho_t$ and $\theta_t$ is called the phase-amplitude method~\cite{Andersson}.
The Wronskian $W_t$ is then given by $W_t=\rho_t^2\dot\theta_t$ by differentiating $\theta_t$ with respect to time $t$.
From this Wronskian represented by $\rho_t$ and $\theta_t$, we can derive a general expression of 
the Ermakov equation based on the phase-amplitude method as (see Appendix~\ref{sec:deriv_ermakoveq_phamp})
\begin{align}
\ddot\rho_t+f_t^2\rho_t=\frac{W_t^2}{\rho_t^3},\label{eq:ermakov_pa}
\end{align}
where 
\begin{align}
f_t
\equiv
\sqrt{\dot\theta_t^2-\frac{3}{4}\frac{\ddot\theta_t^2}{\dot\theta_t^2}+\frac{1}{2}\frac{\dddot\theta_t}{\dot\theta_t}}.
\end{align}
Since $\rho_t$ obeys Eq.~(\ref{eq:Ermakov_eq}), by comparing the two expressions, 
we find $\tilde\varOmega_t=f_t$ from Eq.~(\ref{eq:omega_tilde}).
If $\dot\omega_{t_0}=0$, we can self-evidently identify the time differential of the phase function $\theta_t$ with the frequency $\omega_t$ as 
\begin{eqnarray}
\dot\theta_t=\omega_t\quad ;\,\dot\omega_{t_0}=0.
\label{eq:theta_dot_omega}
\end{eqnarray}
The Wronskian $W_t$ is thus given by
\begin{align}
W_t=\rho_t^2\omega_t\quad ;\,\dot\omega_{t_0}=0.
\label{eq:wronskian_rho_omega}
\end{align}
Noting $W_t=1$, we obtain the explicit solution of the Ermakov equation (Eq.~(\ref{eq:Ermakov_eq})) as
\begin{eqnarray}
\rho_t
=\frac{1}{\sqrt{\omega_t}}
\quad ;\,\dot\omega_{t_0}=0.
\label{eq:rho_solution}
\end{eqnarray}
Here, the condition of $\dot\omega_{t_0}=0$ in Eqs.~(\ref{eq:theta_dot_omega})--(\ref{eq:rho_solution}) is necessary 
for the following reason. 
From the definition of $\rho_t$ in Eq.~(\ref{eq:def_rho}), we find
$\rho_{t_0}=\sqrt{\frac{\varOmega_{t_0}^2\mu_{t_0}^2+\nu_{t_0}^2}{\varOmega_{t_0}}}=\frac{1}{\sqrt{\varOmega_{t_0}}}$
by using Eqs.~(\ref{eq:cpo_mu}) and (\ref{eq:cpo_nu}).
For this expression to be consistent with $\frac{1}{\sqrt{\omega_{t_0}}}$, we must require $\dot{\omega}_{t_0}=0$.
By substituting Eq.~(\ref{eq:rho_solution}) into Eqs.~(\ref{eq:mu_ermakov_formal}) and (\ref{eq:nu_ermakov_formal}), we obtain Eqs.~(\ref{eq:mu_simple_sol}) and (\ref{eq:nu_simple_sol}), respectively.

We next derive Eqs.~(\ref{eq:energy_mu}) and (\ref{eq:energy_nu}).
We define the Ermakov-Lewis invariant~\cite{Lewis} as
\begin{empheq}[left={I_t\equiv\empheqlbrace}]{align}
&\frac{\varOmega_{t_0}}{2}\biggl\{(\dot\rho_t\mu_t-\rho_t\dot\mu_t)^2+W_t^{(\mu)2}\frac{\mu_t^2}{\rho_t^2}\biggr\}
\equiv
I_t^{(\mu)},
\label{eq:ELinv_mu}\\
&\frac{1}{2\varOmega_{t_0}}\biggl\{(\dot\rho_t\nu_t-\rho_t\dot\nu_t)^2+W_t^{(\nu)2}\frac{\nu_t^2}{\rho_t^2}\biggr\}
\equiv
I_t^{(\nu)}.
\label{eq:ELinv_nu}
\end{empheq}
This invariant is indeed a dynamical invariant for the CPO 
obeying equations with linearly independent solutions such as Eqs.~(\ref{eq:cpo_mu}) and (\ref{eq:cpo_nu}).
The Ermakov-Lewis invariant can be shown to be equivalent to one-half of the square of the Wronskian from Eqs.~(\ref{eq:wronskian_mu}) and (\ref{eq:wronskian_nu}) as~\cite{Guasti09}
\begin{align}
I_t=I_t^{(\mu)}=I_t^{(\nu)}=\frac{W_t^2}{2}.
\label{eq:EL_inv_wron}
\end{align}
Then, from Eqs.~(\ref{eq:wronskian_rho_omega}) and (\ref{eq:ELinv_mu})--(\ref{eq:EL_inv_wron}), 
we rewrite the Wronskian $W_t$ in terms of the coordinate variables $\mu_t$ and $\nu_t$ and frequency $\omega_t$ as (see Appendix~\ref{sec:deriv_wr})
\begin{align}
W_t^{(\mu)}
&=
\frac{2\omega_{t_0}}{\omega_t}\biggl\{\mathcal{E}_t^{(\mu)}+\biggl(\dot\mu_t+\frac{\mu_t}{2}\frac{\dot\omega_t}{\omega_t}\biggr)\frac{\mu_t}{2}\frac{\dot\omega_t}{\omega_t}\biggr\}&
&;\,\dot\omega_{t_0}=0,
\label{eq:wronskian_mu_omega}\\
W_t^{(\nu)}
&=
\frac{2}{\omega_t\omega_{t_0}}\biggl\{\mathcal{E}_t^{(\nu)}+\biggl(\dot\nu_t+\frac{\nu_t}{2}\frac{\dot\omega_t}{\omega_t}\biggr)\frac{\nu_t}{2}\frac{\dot\omega_t}{\omega_t}\biggr\}&
&;\,\dot\omega_{t_0}=0.
\label{eq:wronskian_nu_omega}
\end{align}
By noting $W_t=W_t^{(\mu)}=W_t^{(\nu)}=1$, we obtain Eqs.~(\ref{eq:energy_mu}) and (\ref{eq:energy_nu}) from Eqs.~(\ref{eq:wronskian_mu_omega}) and (\ref{eq:wronskian_nu_omega}), respectively.

\subsection{Example}
We visually illustrate the effect of the $\hat{H}^{\rm cd}_t$ term of the QPO on the corresponding CPO on the classical phase space. 
We consider a specific case where the frequency of the QPO is given by a cubic function of time $t\in[t_0,t_{\rm f}]$ as
\begin{align}
\omega_t
=
\omega_0+(\omega_{\rm f}-\omega_0)\biggl\{1+2\frac{(t_{\rm f}-t_0)(t_{\rm f}-t)}{t_0^2+t_{\rm f}^2}\biggr\}\biggl(\frac{t-t_0}{t_{\rm f}-t_0}\biggr)^2,
\label{eq:cubic}
\end{align}
which satisfies $\omega_{t_0}=\omega_0$, $\omega_{t_{\rm f}}=\omega_{\rm f}$, and $\dot\omega_{t_0}=\dot\omega_{t_{\rm f}}=0$.
Here, we consider three cases with different final times $t_{\rm f}=0.2,\, 0.5,\, 2.0$ and set $t_0=0$ and $(\omega_0,\omega_{\rm f})=(2,4)$.

In Fig.~\ref{fig:phase_space}, we show some phase-space trajectories of the CPO given by Eqs.~(\ref{eq:cpo_mu}) and (\ref{eq:cpo_nu}) 
and by Eqs.~(\ref{eq:cpo_mu_omega}) and (\ref{eq:cpo_nu_omega}), the corresponding Hamiltonians of which as the QPO 
are $\hat{H}^{\rm TT}_t$ and $\hat{H}^{\rm ad}_t$, respectively.
For every final time $t_{\rm f}$, we can find that the final points of the trajectories given by Eqs.~(\ref{eq:cpo_mu}) and (\ref{eq:cpo_nu}) are always on the same targeted energy shells $\mathcal{E}^{(\mu)}_{t_{\rm f}}=\frac{\omega_{t_{\rm f}}}{2\omega_{t_0}}=\frac{\omega_{\rm f}}{2\omega_0}$
and
$\mathcal{E}^{(\nu)}_{t_{\rm f}}=\frac{\omega_{t_0}\omega_{t_{\rm f}}}{2}=\frac{\omega_0\omega_{\rm f}}{2}$ with the aid of the $\hat{H}^{\rm cd}$ term, 
implying the success of the TT algorithm.
On the other hand, $E_{t_{\rm f}}^{(\mu)}$ and $E_{t_{\rm f}}^{(\nu)}$ at the final points of the trajectories given by Eqs.~(\ref{eq:cpo_mu_omega}) and (\ref{eq:cpo_nu_omega}), respectively, vary depending on $t_{\rm f}$, 
failing to achieve the same final state as that of the adiabatic processes unless a sufficiently large $t_{\rm f}$ is taken.

In Fig.~\ref{fig:Qtt_vs_Q*}, we show the time dependence of the parameters $Q^{\rm TT}_t$ and $Q^*_t$ together with the quantities $\frac{\mathcal E^{(k)}_t}{\varOmega_t}$ and $\frac{E^{(k)}_t}{\omega_t}$ ($k=\mu, \nu$) in the insets, which are obtained using the data of the trajectories in Fig.~\ref{fig:phase_space}.
We can find that $\frac{\mathcal E^{(k)}_t}{\varOmega_t}=J^{(k)}$ holds at the final points at $t=t_{\rm f}$ as expected.
That is, $Q^{\rm TT}_t$ is unity at every final time $t_{\rm f}$ we chose, but it is not so for the intermediate times .
Without the $\hat{H}^{\rm cd}$ term, the areas enclosed by the trajectories were well defined and $\frac{E^{(k)}_t}{\omega_t}=J^{(k)}$ holds for an arbitrary time $t$ only if a sufficiently large $t_{\rm f}$ is taken.

\begin{widetext}

\begin{figure}[t]
\begin{minipage}[htbp]{0.39\hsize}
\begin{center}
\subfigure{%
\includegraphics[width=0.9\linewidth]{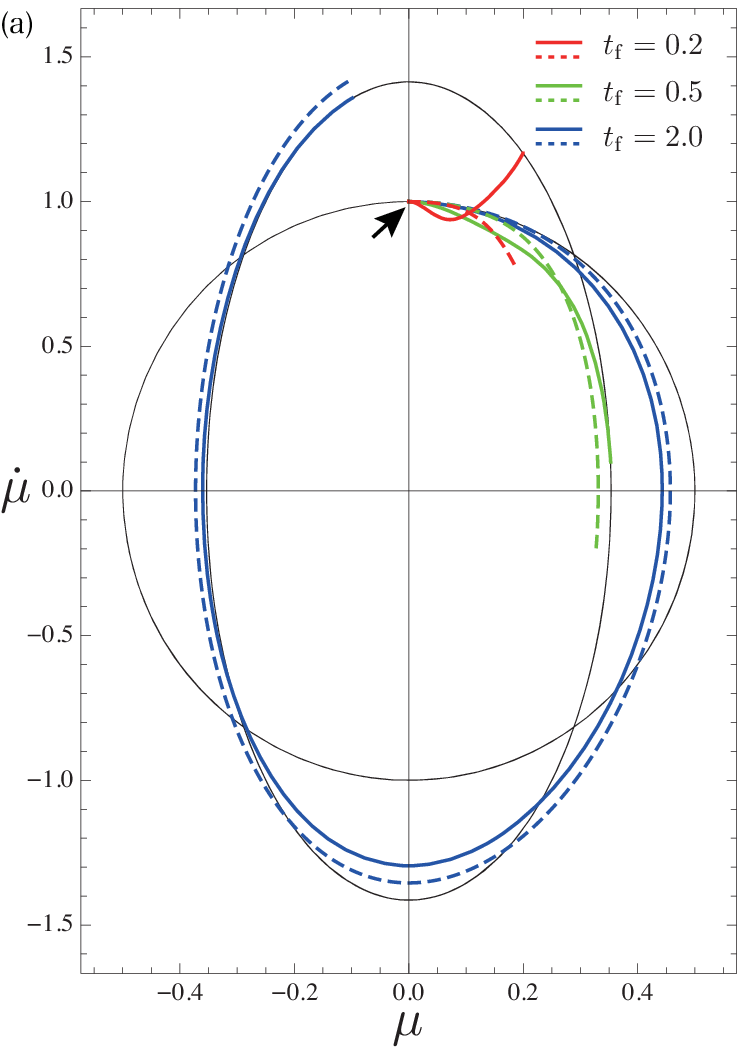}}
\label{ps_mu}\end{center}
\end{minipage}\begin{minipage}[htbp]{0.39\hsize}
\begin{center}
\subfigure{%
\includegraphics[width=0.9\linewidth]{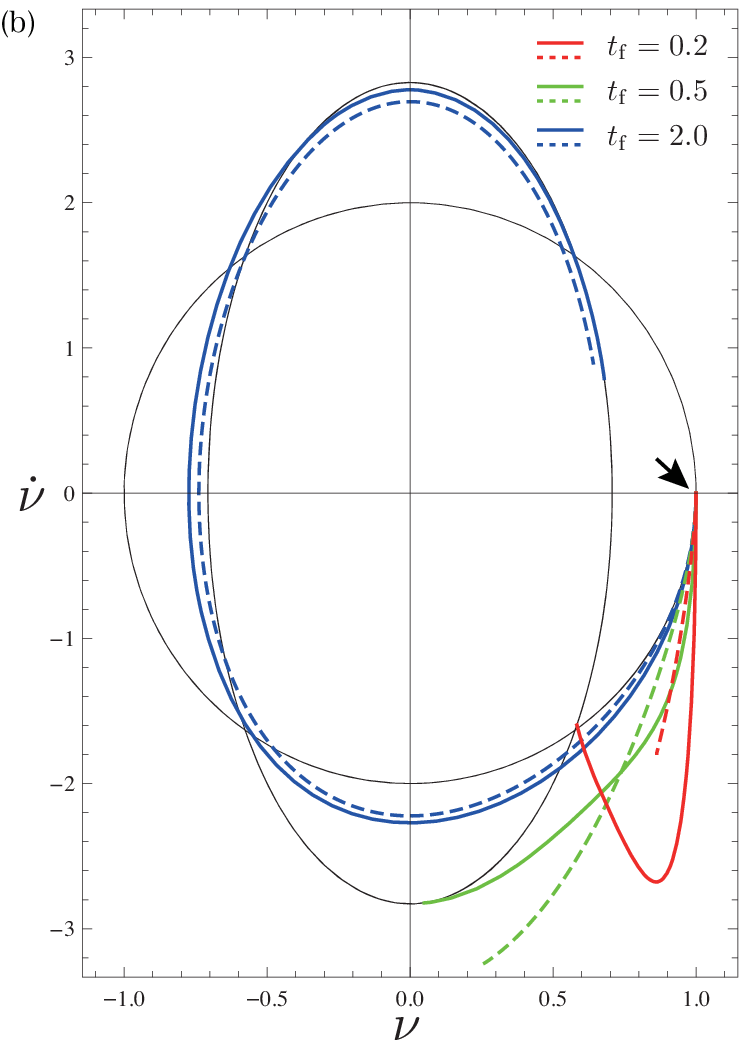}}
\label{ps_nu}
\end{center}
\end{minipage}
\vspace{-1em}
\caption{%
Trajectories of the CPO: (a) trajectories of $\mu_t$ obeying Eqs.~(\ref{eq:cpo_mu}) (solid lines) and (\ref{eq:cpo_mu_omega}) (dashed lines) and (b) trajectories of $\nu_t$ obeying Eqs.~(\ref{eq:cpo_nu}) (solid lines) and (\ref{eq:cpo_nu_omega}) (dashed lines) 
for different final times on the classical phase space.
The arrows indicate the initial points $(0, 1)$ and $(1, 0)$ for $\mu_t$ and $\nu_t$, respectively.
The ellipses with these initial points denote the trajectories of the CPO with the initial ``energies'' $\mathcal E^{(\mu)}_{t_0}=\frac{1}{2}$ and $\mathcal E^{(\nu)}_{t_0}=\frac{\omega_0^2}{2}=2.0$.
The other ellipses denote the trajectories of the CPO with the final ``energies''
$\mathcal E^{(\mu)}_{t_{\rm f}}=\frac{\omega_{\rm f}}{2\omega_0}=1.0$ and $\mathcal E^{(\nu)}_{t_{\rm f}}=\frac{\omega_0\omega_{\rm f}}{2}=4.0$.
}
\label{fig:phase_space}
\end{figure}

\begin{figure}[t]
\begin{center}
\includegraphics[width=0.7\linewidth]{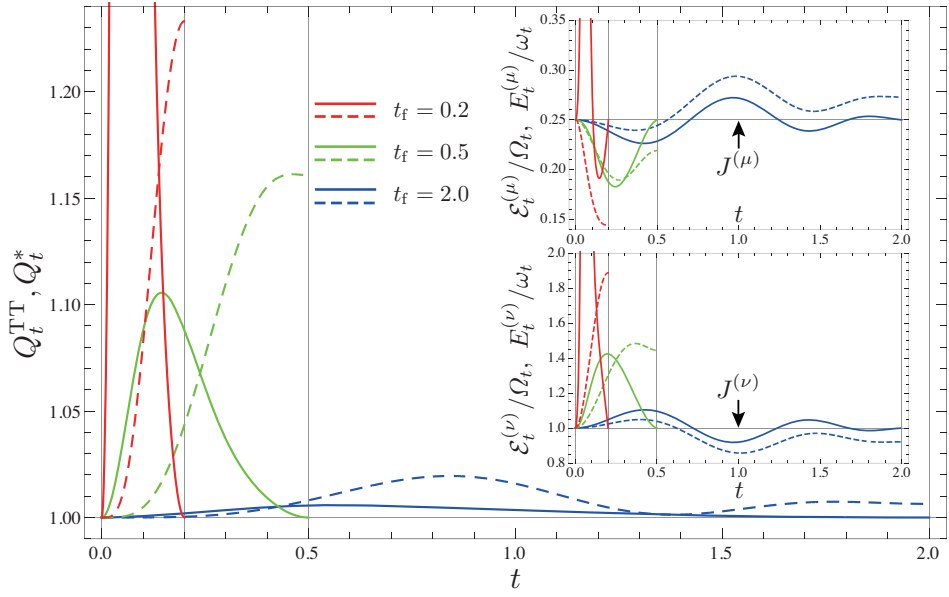}
\label{ps_mu}
\end{center}
\vspace{-1em}
\caption{%
Comparison of $Q^{\rm TT}_t$ (solid line) and $Q^*_t$ (dashed line) as functions of time $t$.
The insets show the comparison of $\frac{\mathcal E^{(k)}_t}{\varOmega_t}$ (solid line) and $\frac{E^{(k)}_t}{\omega_t}$ (dashed line) ($k=\mu, \nu$) as functions of time $t$, where $J^{(\mu)}=\frac{1}{2\omega_0}=0.25$ and $J^{(\nu)}=\frac{\omega_0}{2}=1.0$.
}
\label{fig:Qtt_vs_Q*}
\end{figure}

\begin{figure}[htbp]
\begin{minipage}[htbp]{0.5\hsize}
\begin{center}
\subfigure{%
\includegraphics[width=1.0\linewidth]{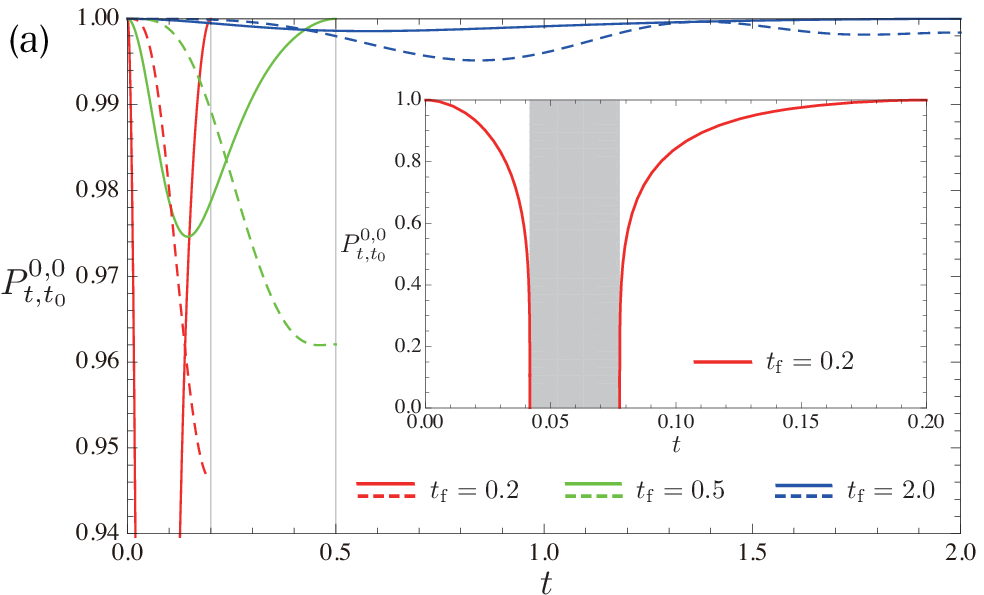}}
\label{fig:pmn_even}\end{center}
\end{minipage}\begin{minipage}[htbp]{0.5\hsize}
\begin{center}
\subfigure{%
\includegraphics[width=1.0\linewidth]{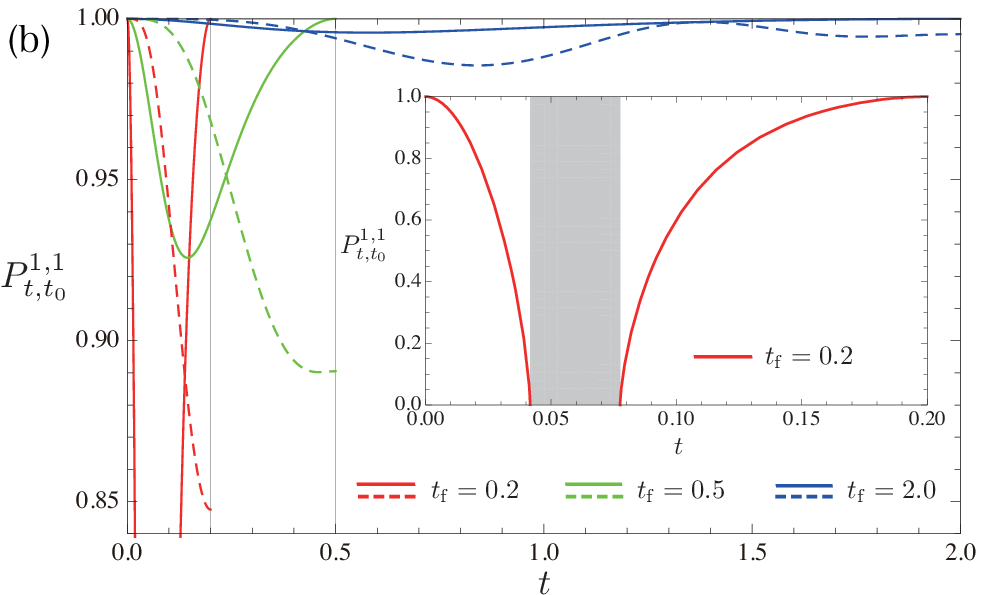}}
\label{fig:pmn_odd}
\end{center}
\end{minipage}
\vspace{-1em}
\caption{%
Transition probabilities between (a) even and (b) odd quantum-number states, $n,m=0$ and $n,m=1$, respectively, as functions of time $t$.
The solid (dashed) lines represent the cases with (without) the counterdiabatic term.
The gray-shaded regions in the insets represent the time intervals in which the transition probabilities are ill-defined for the case of $t_{\rm f}=0.2$.
}
\label{fig:pmn}
\end{figure}

\end{widetext}

In Fig.~\ref{fig:pmn}, we show two transition probabilities $P_{t,t_0}^{0,0}$ and $P_{t,t_0}^{1,1}$ 
obtained using the time evolution of $Q^{\rm TT}_t$ and $Q^*_t$ in Fig.~\ref{fig:Qtt_vs_Q*}.
By a selection rule~\cite{Husimi}, only transitions between even or odd quantum-number states are allowed
(see Appendix~\ref{sec:exact_pmn}).

We note that, in the fastest case of $t_{\rm f}=0.2$ in Fig.~\ref{fig:Qtt_vs_Q*}, 
$\varOmega_t$ temporarily attains an imaginary value after
$Q^{\rm TT}_t$ shows diverging behavior as $\varOmega_t \to 0$. 
Because $\hat{H}_{\rm TT}$ in Eq.~(\ref{eq:hamiltonian _tt_r}) has a continuous energy spectrum in this time interval, 
the transition probabilities in Eq.~(\ref{eq:trans_prob}) using the discrete energy spectrum cannot be defined there 
(the gray-shaded regions in the insets of Fig.~\ref{fig:pmn} represent these time intervals).
As the time evolution approaches the final state, however, $\varOmega_t$ recovers its real value.

\section{Concluding remarks}
\label{sec:sum}

Here, we give a few remarks on the main result of the probability generating function of the transitionless QPO given by Eq.~(\ref{eq:pgf_TT}).

First, we note that 
the phase-amplitude method is formally applicable to 
the CPO in Eqs.~(\ref{eq:cpo_mu_omega}) and (\ref{eq:cpo_nu_omega}) 
induced from the usual QPO without the counterdiabatic term; that is, it is formally applicable to the original case considered by Husimi~\cite{Husimi}.
In this case, the Ermakov equation given by Eq.~(\ref{eq:Ermakov_eq}) is replaced with
\begin{align}
\ddot\rho_t+\omega_t^2\rho_t
=
\frac{W_t^2}{\rho_t^3}.
\label{eq:Ermakov_eq_husimi}
\end{align}
With this replacement, the simple identification $\dot{\theta}_t=\omega_t$ in Eq.~(\ref{eq:theta_dot_omega}) is no longer applicable, and 
we cannot obtain a simple solution such as Eq.~(\ref{eq:rho_solution}) for this case in general.
However, in the adiabatic approximation of $\ddot{\rho}_t \simeq 0$, 
the Ermakov equation in Eq.~(\ref{eq:Ermakov_eq_husimi}) has the solution given by Eq.~(\ref{eq:rho_solution}) as an adiabatic solution~\cite{Beau16}: 
\begin{eqnarray}
\rho_t \simeq \frac{1}{\sqrt{\omega_t}}.
\end{eqnarray}
This implies that the Ermakov equation given by Eq.~(\ref{eq:Ermakov_eq}) for the transitionless QPO 
can have the adiabatic solution of the original Ermakov equation given by Eq.~(\ref{eq:Ermakov_eq_husimi}) for the usual QPO
as the exact solution, as the Schr\"odinger equation with the TT Hamiltonian 
can have the adiabatic solution of the original Schr\"odinger equation 
with the adiabatic Hamiltonian as the exact solution.
This is an interpretation of the TT algorithm applied to the QPO 
based on the transition probability generating function approach via Husimi's method.

Second,
Ref.~\cite{Beau16} recently introduced the ratio of the nonadiabatic mean energy to the adiabatic one 
as a nonadiabatic factor to characterize the shortcuts to adiabaticity of a quantum heat engine, 
which is represented by a scale factor satisfying the Ermakov equation.
This nonadiabatic factor has been shown to be equivalent to the original Husimi's measure of adiabaticity under scale-invariant dynamics~\cite{Jaramillo}.
In contrast, our parameter is derived from the direct calculation of the probability generating function given by Eq.~(\ref{eq:pgf_TT}) and includes the nonadiabatic factor as a special case.
By substituting Eqs.~(\ref{eq:mu_ermakov_formal}) and (\ref{eq:nu_ermakov_formal}) into Eq.~(\ref{eq:Q_TT_general}) and using Eq.~(\ref{eq:rho_solution}), we can express $Q^{\rm TT}_t$
by using $\rho_t$ as
\begin{align}
Q^{\rm TT}_t
&=
\frac{1}{2\varOmega_t}\biggl(\dot\rho_t^2+\varOmega_t^2\rho_t^2+\frac{W_t^2}{\rho_t^2}\biggr)
\quad ;\,\dot\omega_{t_0}=0.
\label{eq:Q_TT_rho}
\end{align}
This expression agrees with the nonadiabatic factor in Ref.~\cite{Beau16} if $\varOmega_t$ is replaced with $\omega_t$.

In this paper, we have studied the transitionless QPO with the TT algorithm based on the transition probability generating function approach.
By applying Husimi's method, we have obtained the propagator of the transitionless QPO with the two linearly independent solutions of the corresponding CPO.
By calculating the probability generating function from the propagator, we have found that it contains a simple time-dependent parameter that 
can characterize the success of the TT algorithm.
The key to obtain this simple parameter was the explicit solutions of the CPO derived based on the phase-amplitude method.
We have illustrated this result by showing the trajectories of the CPO on the classical phase space and the time dependence of our parameter by using a specific form of the frequency.
We hope that the present work provides a new perspective on the shortcuts to adiabaticity.

\acknowledgments
The authors are grateful to K. Nakamura for guiding us to study the present subject and for helpful discussions.
They are also grateful to S. Tanimura for helpful discussions and critical reading of the manuscript, 
and to S. Masuda for helpful discussions.


\appendix

\section{Derivation of energy eigenfunction in Eq.~(\ref{eq:efunc_tt})}
\label{sec:deriv_energy_ef}

From the definition of the vacuum state $\hat{b}_t\ket{0;\varOmega_t}\equiv 0$ and by using Eq.~(\ref{eq:boson_op}), we can easily obtain the normalized energy eigenfunction of the vacuum state for position $x$ as
\begin{align}
\bket{x}{0;\varOmega_t}
=
\biggl(\frac{M\varOmega_t}{\pi\hbar}\biggr)^{1/4}\exp\biggl(-\frac{\zeta_t M\varOmega_t}{2\hbar}x^2\biggr).
\end{align}
\begin{widetext}
\noindent For a function $f(x)$, by using the relation
\begin{align}
\biggl(\sqrt{a}x-\frac{1}{\sqrt{a}}\pdiff{}{x}\biggr)^n f(x)
=
(-1)^n e^{ax^2/2}\frac{1}{a^{n/2}}\pdiff{^n}{x^n}\big(e^{-ax^2/2}f(x)\big),
\label{eq:form1}
\end{align}
we can also obtain the normalized energy eigenfunction of the $n$-th excited state in Eq.~(\ref{eq:efunc_tt}) as follows:
\begin{align}
\bket{x}{n;\varOmega_t}
&=
\frac{1}{\sqrt{n!}}\bketm{x}{\hat{b}^{\dagger n}_t}{0;\varOmega_t}
=
\frac{\zeta^{*n/2}}{\sqrt{2^n n!}}\biggl(\sqrt{\frac{\zeta^*_t M\varOmega_t}{\hbar}}x-\sqrt{\frac{\hbar}{\zeta^*_t M\varOmega_t}}\pdiff{}{x}\biggr)^n\bket{x}{0;\varOmega_t}
\notag\\
&=
\frac{1}{\sqrt{2^n n!}}(-1)^n\exp\biggl(\frac{\zeta^*_t M\varOmega_t}{2\hbar}x^2\biggr)\biggl(\frac{\hbar}{M\varOmega_t}\biggr)^{n/2}
\pdiff{^n}{x^n}\biggl\{\exp\biggl(-\frac{\zeta^*_t M\varOmega_t}{2\hbar}x^2\biggr)\bket{x}{0;\varOmega_t}\biggr\}
\notag\\
&=
\frac{1}{\sqrt{2^n n!}}\biggl(\frac{M\varOmega_t}{\pi\hbar}\biggr)^{1/4}(-1)^n\exp\biggl(\frac{M\varOmega_t}{\hbar}x^2\biggr)\biggl(\frac{\hbar}{M\varOmega_t}\biggr)^{n/2}
\biggl\{\pdiff{^n}{x^n}\exp\biggl(-\frac{M\varOmega_t}{\hbar}x^2\biggr)\biggr\}\exp\biggl(-\frac{\zeta_t M\varOmega_t}{2\hbar}x^2\biggr)
\notag\\
&=
\frac{1}{\sqrt{2^n n!}}\biggl(\frac{M\varOmega_t}{\pi\hbar}\biggr)^{1/4}{\rm H}_n\biggl(\sqrt{\frac{M\varOmega_t}{\hbar}}x\biggr)\exp\biggl(-\frac{\zeta_t M\varOmega_t}{2\hbar}x^2\biggr),
\label{eq:efunc_tt_SM}
\end{align}
\end{widetext}
where the $n$-th-degree Hermite polynomial ${\rm H}_n$ is defined as
\begin{align}
{\rm H}_n(x)
\equiv
(-1)^n e^{x^2}\diff{^n}{x^n}e^{-x^2}.
\label{eq:hermite_poly}
\end{align}


\section{Derivation of propagator in Eq.~(\ref{eq:gf_tt})}
\label{sec:deriv_prop}

Based on Husimi's method~\cite{Husimi}, we derive Eq.~(\ref{eq:gf_tt}).
For the TT Hamiltonian of the QPO in Eq.~(\ref{eq:hamiltonian_tt}), the $x$-representation of the wave function
\begin{align}
\bket{x}{\psi_n(t)}
=
\int_\mathbb{R}\mathrm{d}x_0U_{t,t_0}(x|x_0)\bket{x_0}{\psi_n(t_0)}
\label{eq:wave_func}
\end{align}
satisfies the Schr\"oginger equation
\begin{align}
{\rm i}\hbar\pdiff{}{t}\bket{x}{\psi_n(t)}
=
\bketm{x}{\hat{H}^\text{TT}_t}{\psi_n(t)},
\label{eq:Schr_prop}
\end{align}
where $U_{t,t_0}(x|x_0)$ is the propagator.
Here, we assume the following Gaussian form of the propagator as the specific anstaz~\cite{Husimi,Suazo}: 
\begin{align}
U_{t,t_0}(x|x_0)
=
\sqrt{\frac{M}{2\pi{\rm i}\hbar\mu_t}}e^{{\rm i}(\alpha_t x^2+\beta_t xx_0+\gamma_t x_0^2)/\hbar},
\label{eq:gf_anstaz}
\end{align}
where the coefficients $\mu_t$, $\alpha_t$, $\beta_t$, and $\gamma_t$ are time-dependent real-valued functions.
By substituting Eq.~(\ref{eq:wave_func}) with Eq.~(\ref{eq:gf_anstaz}) into the Schr\"oginger equation given by Eq.~(\ref{eq:Schr_prop}), we find that 
four coupled ordinary differential equations for the coefficients $\mu_t$, $\alpha_t$, $\beta_t$, and $\gamma_t$ follow:
\begin{empheq}[left=\empheqlbrace]{align}
&\alpha_t
=
\frac{M}{2}\biggl(\frac{\dot\mu_t}{\mu_t}+\frac{1}{2}\frac{\dot\omega_t}{\omega_t}\biggr),
\label{eq:system_alpha&mu}\\
&\dot\alpha_t+\frac{2}{M}\alpha^2_t-\frac{\dot\omega_t}{\omega_t}\alpha_t+\frac{M}{2}\omega^2_t
=
0,
\label{eq:system_alpha}\\
&\dot\beta_t+\biggl(\frac{2}{M}\alpha_t-\frac{1}{2}\frac{\dot{\omega}_t}{\omega_t}\biggr)\beta_t
=
0,
\label{eq:system_beta}\\
&\dot\gamma_t+\frac{\beta^2_t}{2M}
=
0.
\label{eq:system_gamma}
\end{empheq}
By substituting Eq.~(\ref{eq:system_alpha&mu}) into Eq.~(\ref{eq:system_alpha}), we obtain
\begin{align}
\ddot{\mu}_t+\tilde{\varOmega}_t^2 \mu_t=0.
\label{eq:system_mu}
\end{align}
By substituting Eq.~(\ref{eq:system_alpha&mu}) into Eq.~(\ref{eq:system_beta}) and by solving Eq.~(\ref{eq:system_beta}) with respect to $\beta_t$, we have 
\begin{align}
\beta_t&=\frac{C_1}{\mu_t},
\label{eq:beta_C1}
\end{align}
with the integral constant $C_1$.
Next, by substituting Eq.~(\ref{eq:beta_C1}) into Eq.~(\ref{eq:system_gamma}) and by solving Eq.~(\ref{eq:system_gamma}) with respect to $\gamma_t$, we also have
\begin{align}
\gamma_t
=
-\frac{C_1^2}{2M}\int^t_{t_0}\frac{\mathrm{d}t'}{\mu^2_{t'}}+C_2,
\label{eq:gamma_C1C2}
\end{align}
with the integral constant $C_2$. From Eqs.~(\ref{eq:system_alpha&mu}), (\ref{eq:beta_C1}), and (\ref{eq:gamma_C1C2}), the dynamics of $\alpha_t$, $\beta_t$, and $\gamma_t$ can be determined 
by using the solution of $\mu_t$ satisfying Eq.~(\ref{eq:system_mu}).

We now determine the initial condition of $\mu_t$ in Eq.~(\ref{eq:system_mu}) and the integral constants $C_1$ and $C_2$ according to the following argument:
to represent the wave function $\bket{x}{\psi_n(t)}$ from an arbitrary initial wave function $\bket{x_0}{\psi_n(t_0)}$, 
the propagator $U_{t,t_0}(x|x_0)$ in Eq.~(\ref{eq:gf_anstaz}) needs to satisfy $\lim_{t\to t_0+0}U_{t,t_0}(x|x_0)=\delta (x-x_0)$.
Therefore, it is natural to assume the following asymptotic form of the propagator :
\begin{align}
&U_{t,t_0}(x|x_0)|_{t\simeq t_0}
\notag\\
&\simeq
\sqrt{\frac{M}{2\pi{\rm i}\hbar(t-t_0)}}\exp\biggl[\frac{{\rm i}M}{2\hbar}\frac{(x-x_0)^2}{t-t_0}\biggr]e^{{\rm i}F(x,x_0)/\hbar},\label{eq:gf_anstaz_asym_s}
\end{align}
where $F(x,x_0)$ is a function that satisfies $F(x_0,x_0)=0$.
Through a Taylor expansion of $\mu_t$ with respect to $t$ around $t_0$ and from Eqs.~(\ref{eq:gf_anstaz}) and (\ref{eq:gf_anstaz_asym_s}), we have
\begin{align}
\mu_t |_{t\simeq t_0}
&=
\mu_{t_0}+\dot{\mu}_{t_0}(t-t_0)+\mathcal{O}((t-t_0)^2)
\notag\\
&=
t-t_0+\mathcal{O}((t-t_0)^2),
\label{eq:asym_mu}
\end{align}
from which we can determine $\mu_{t_0}$ and $\dot{\mu}_{t_0}$ as
\begin{align}
\mu_{t_0}=0, \quad \dot\mu_{t_0}=1,
\label{eq:mu_initial}
\end{align}
as the initial condition of Eq.~(\ref{eq:system_mu}).
By using $\ddot\mu_{t_0}=0$ obtained from Eqs.~(\ref{eq:system_mu}) and (\ref{eq:mu_initial}), we modify Eq.~(\ref{eq:asym_mu}) as
\begin{align}
\mu_t|_{t\simeq t_0}=t-t_0+\mathcal{O}((t-t_0)^3).
\label{eq:asym_mu_strict}
\end{align}
From Eq.~(\ref{eq:system_alpha&mu}), we can find
\begin{align}
\alpha_t|_{t\simeq t_0}
=
\frac{M}{2}\biggl(\frac{1}{t-t_0}+\frac{1}{2}\frac{\dot\omega_{t_0}}{\omega_{t_0}}\biggr)+\mathcal{O}(t-t_0).
\label{eq:asym_alpha}
\end{align}
From Eqs.~(\ref{eq:beta_C1}) and (\ref{eq:asym_mu_strict}), we then find
\begin{align}
\beta_t|_{t\simeq t_0}
=
\frac{C_1}{t-t_0}+\mathcal{O}(t-t_0).
\label{eq:asym_beta_C1}
\end{align}
To determine the asymptotic form of $\gamma_t$ in Eq.~(\ref{eq:gamma_C1C2}),
we introduce a solution $\nu_t$ as a linearly independent solution of $\mu_t$, which satisfies the same equation as 
Eq.~(\ref{eq:system_mu}) but with a different initial condition:
\begin{eqnarray}
\nu_{t_0}=1, \quad  \dot{\nu}_{t_0}=0,\label{eq:nu_initial}
\end{eqnarray}
where the Wronskian $W_t=\dot{\mu}_t \nu_t-\mu_t\dot{\nu}$ is unity as in Eq.~(\ref{eq:wronskian}).
From this Wronskian, we obtain
\begin{align}
\frac{\nu_t}{\mu_t}=-\int_{t_0}^t\frac{\mathrm{d}t'}{\mu^2_{t'}}.
\label{eq:nu_over_mu}
\end{align}
From Eqs.~(\ref{eq:gamma_C1C2}) and (\ref{eq:nu_over_mu}), we have
\begin{align}
\gamma_t
=
\frac{C_1^2}{2M}\frac{\nu_t}{\mu_t}+C_2.
\label{eq:gamma_C1C2_nu}
\end{align}
From Eqs.~(\ref{eq:asym_mu}), (\ref{eq:mu_initial}), and (\ref{eq:nu_initial}), we obtain the asymptotic form of $\gamma_t$ as
\begin{align}
\gamma_t|_{t\simeq t_0}
=
\frac{C_1^2}{2M}\frac{1}{t-t_0}+C_2+\mathcal{O}(t-t_0).
\label{eq:asym_gamma_C1C2_nu}
\end{align}
By substituting Eqs.~(\ref{eq:asym_mu}), (\ref{eq:asym_alpha}), (\ref{eq:asym_beta_C1}), and (\ref{eq:asym_gamma_C1C2_nu}) into Eq.~(\ref{eq:gf_anstaz}), we have the following asymptotic form of the propagator:
\begin{align}
&U_{t,t_0}(x|x_0)|_{t\simeq t_0}
\notag\\
&=
\sqrt{\frac{M}{2\pi {\rm i}\hbar(t-t_0)}}
\exp\biggl[\frac{\rm i}{\hbar}\biggl\{\frac{M}{2}\biggl(\frac{1}{t-t_0}+\frac{1}{2}\frac{\dot\omega_{t_0}}{\omega_{t_0}}\biggr)x^2
\notag\\
&\phantom{=}
+\frac{C_1}{t-t_0}xx_0+\biggl(\frac{C_1^2}{2M}\frac{1}{t-t_0}+C_2\biggr)x_0^2\biggr\}+\mathcal{O}(t-t_0)\biggr]
\notag\\
&=
\sqrt{\frac{M}{2\pi {\rm i}\hbar(t-t_0)}}
\exp\biggl[\frac{{\rm i}M}{2\hbar}\frac{(x-x_0)^2}{t-t_0}\biggr]e^{{\rm i}F(x,x_0)/\hbar},
\label{eq:prop_F}
\end{align}
where the function $F(x,x_0)$ is
\begin{align}
F(x,x_0)
&=
\frac{M}{4}\frac{\dot\omega_{t_0}}{\omega_{t_0}}x^2+\frac{C_1+M}{t-t_0}xx_0
\notag\\
&\phantom{=}
+\biggl(\frac{C_1^2-M^2}{2M}\frac{1}{t-t_0}+C_2\biggr)x_0^2+\mathcal{O}(t-t_0).
\end{align}
Since $F(x_0,x_0)=0$ is required in the limit of $t\to t_0+0$, we must set $C_1=-M$ and $C_2=-\frac{M}{4}\frac{\dot\omega_{t_0}}{\omega_{t_0}}$.
We then obtain
\begin{align}
\beta_t=-\frac{M}{\mu_t},
\quad
\gamma_t=\frac{M}{2}\biggl(\frac{\nu_t}{\mu_t}-\frac{1}{2}\frac{\dot\omega_{t_0}}{\omega_{t_0}}\biggr).
\label{eq:beta_gamma}
\end{align}
By substituting Eqs.~(\ref{eq:system_alpha&mu}) and (\ref{eq:beta_gamma}) into Eq.~(\ref{eq:gf_anstaz}), we finally obtain the propagator given by Eq.~(\ref{eq:gf_tt}).

\begin{widetext}

\section{Derivation of probability generating function in Eq.~(\ref{eq:pgf_TT})}
\label{sec:deriv_pgf}

By using Mehler's formula and the energy eigenfunction given by Eq.~(\ref{eq:efunc_tt_SM}) (Eq.~(\ref{eq:efunc_tt})), we obtain the following relation:
\begin{align}
\sum_{n=0}^\infty z^n\bket{n;\varOmega_t}{x}\bket{y}{n;\varOmega_t}
=
\sqrt{\frac{M\varOmega_t}{\pi\hbar(1-z^2)}}\exp\biggl[-\frac{M\varOmega_t}{2\hbar}\frac{(1+z^2)(x^2+y^2)-4zxy}{1-z^2}-\frac{{\rm i}M}{4\hbar}\frac{\dot\omega_t}{\omega_t}(x^2-y^2)\biggr].
\label{eq:mehler}
\end{align}
By using Eq.~(\ref{eq:mehler}), we can calculate the probability generating function as
\begin{align}
\mathcal{P}^{u,v}_{t,t_0}
&=
\sum_{n,m=0}^\infty u^n v^m P^{m,n}_{t,t_0}
=
\sum_{n,m=0}^\infty u^n v^m\biggl|\iint_{\mathbb{R}^2}{\rm d}x{\rm d}x_0\bket{m;\varOmega_t}{x}U_{t,t_0}(x|x_0)\bket{x_0}{n;\varOmega_{t_0}}\biggr|^2
\notag\\
&=
\iiiint_{\mathbb{R}^4}{\rm d}x{\rm d}x_0{\rm d}x'{\rm d}x'_0U_{t,t_0}^*(x|x_0)U_{t,t_0}(x'|x'_0)\sum_{m=0}^\infty v^m\bket{m;\varOmega_t}{x}\bket{x'}{m;\varOmega_t}\sum_{n=0}^\infty u^n\bket{n;\varOmega_{t_0}}{x_0}\bket{x'_0}{n;\varOmega_{t_0}}
\notag\\
&=
\frac{2}{\mu_t}\biggl(\frac{M}{2\pi \hbar}\biggr)^2\sqrt{\frac{\varOmega_t\varOmega_{t_0}}{(1-u^2)(1-v^2)}}\int_{\mathbb R^4}{\rm d}\vec x\exp\biggl(-\frac{M}{2\hbar}\vec x\cdot A\vec x\biggr)
\notag\\
&=
\frac{2}{\mu_t}\sqrt{\frac{\varOmega_t\varOmega_{t_0}}{(1-u^2)(1-v^2)\det A}},
\label{eq:pgf_TT_derivation}
\end{align}
where we defined
\begin{align}
\vec x
\equiv
\begin{pmatrix}
\\[-7.5pt]
x\\[3pt]
x_0\\[3pt]
x'\\[3pt]
x'_0\\[4.5pt]
\end{pmatrix},
\quad
A
\equiv 
\begin{pmatrix}
\frac{1+v^2}{1-v^2}\varOmega_t+{\rm i}\bigl(\frac{\dot\mu_t}{\mu_t}+\frac{\dot\omega_t}{\omega_t}\bigr)&-\frac{\rm i}{\mu_t}&-\frac{2v}{1-v^2}\varOmega_t&0\\
-\frac{\rm i}{\mu_t}&\frac{1+u^2}{1-u^2}\varOmega_{t_0}+{\rm i}\frac{\nu_t}{\mu_t}&0&-\frac{2u}{1-u^2}\varOmega_{t_0}\\
-\frac{2v}{1-v^2}\varOmega_t&0&\frac{1+v^2}{1-v^2}\varOmega_t-{\rm i}\bigl(\frac{\dot\mu_t}{\mu_t}+\frac{\dot\omega_t}{\omega_t}\bigr)&\frac{\rm i}{\mu_t}\\
0&-\frac{2u}{1-u^2}\varOmega_{t_0}&\frac{\rm i}{\mu_t}&\frac{1+u^2}{1-u^2}\varOmega_{t_0}-{\rm i}\frac{\nu_t}{\mu_t}
\end{pmatrix},
\end{align}
and used the following formula of the Gaussian integral:
\begin{align}
\int_{\mathbb R^n}{\rm d}\vec x e^{-a\vec x\cdot A\vec x}
=
\sqrt{\frac{(\pi/a)^n}{\det A}},
\end{align}
provided $a>0$, $\vec x\in\mathbb R^n$, and the $n$-by-$n$ matrix $A$ is symmetric.
By using the Wronskian given by Eq.~(\ref{eq:wronskian}), we obtain
\begin{align}
\det A
=
\frac{1}{\mu^2_t}\frac{2\varOmega_t\varOmega_{t_0}}{(1-u^2)(1-v^2)}\bigl\{Q^{\rm TT}_t(1-u^2)(1-v^2)+(1+u^2)(1+v^2)-4uv\bigr\}.\label{eq:detA}
\end{align}
Here, $Q^{\rm TT}_t$ is given as Eq.~(\ref{eq:Q_TT_general}).
Then, we finally obtain Eq.~(\ref{eq:pgf_TT}) by substituting Eq.~(\ref{eq:detA}) into Eq.~(\ref{eq:pgf_TT_derivation}).
\end{widetext}


\section{Derivation of Ermakov equation in Eq.~(\ref{eq:Ermakov_eq}) from Wronskian}\label{sec:deriv_ermakoveq}
\newpage
Here, we derive the Ermakov equation in Eq.~(\ref{eq:Ermakov_eq})~\cite{Ermakov,Pinney50,Leach08}.
By differentiating Eq.~(\ref{eq:wronskian_mu}) with respect to time $t$ and using Eqs.~(\ref{eq:cpo_mu}) and (\ref{eq:wronskian}),
we can show the following relation:
\begin{align}
0&=
\diff{W^{(\mu)}_t}{t}
\notag\\
&=
-\frac{\rho_t\mu_t}{\sqrt{\varOmega_{t_0}^{-1}\rho_t^2-\mu_t^2}}\biggl\{\frac{\ddot\rho_t\mu_t-\rho_t\ddot\mu_t}{\mu_t}-\frac{(\dot\rho_t\mu_t-\rho_t\dot\mu_t)^2}{\rho_t(\varOmega_{t_0}^{-1}\rho_t^2-\mu_t^2)}\biggr\}
\notag\\
&=
\frac{\mu_tW_t^{(\mu)}}{\dot\rho_t\mu_t-\rho_t\dot\mu_t}\biggl(\ddot\rho_t+\tilde\varOmega_t^2\rho_t-\frac{W_t^{(\mu)2}}{\rho_t^3}\biggr).
\end{align}
From the above, we have the Ermakov equation of $\rho_t$ for $\mu_t$:
\begin{align}
\ddot\rho_t+\tilde\varOmega_t^2\rho_t
=
\frac{W^{(\mu)2}_t}{\rho_t^3}.
\label{eq:Ermakov_eq_mu}
\end{align}
Similarly, by differentiating Eq.~(\ref{eq:wronskian_nu}) with respect to time $t$ and using Eqs.~(\ref{eq:cpo_nu}) and (\ref{eq:wronskian}), we can obtain 
the following relation:
\begin{align}
0&=
\diff{W^{(\nu)}_t}{t}
\notag\\
&=
\frac{\rho_t\nu_t}{\sqrt{\varOmega_{t_0}\rho_t^2-\nu_t^2}}\biggl\{\frac{\ddot\rho_t\nu_t-\rho_t\ddot\nu_t}{\nu_t}-\frac{(\dot\rho_t\nu_t-\rho_t\dot\nu_t)^2}{\rho_t(\varOmega_{t_0}\rho_t^2-\nu_t^2)}\biggr\}
\notag\\
&=
\frac{\nu_tW_t^{(\nu)}}{\dot\rho_t\nu_t-\rho_t\dot\nu_t}\biggl(\ddot\rho_t+\tilde\varOmega_t^2\rho_t-\frac{W_t^{(\nu)2}}{\rho_t^3}\biggr).
\end{align}
From the above, we have the Ermakov equation of $\rho_t$ for $\nu_t$:
\begin{align}
\ddot\rho_t+\tilde\varOmega_t^2\rho_t
=
\frac{W^{(\nu)2}_t}{\rho_t^3}.
\label{eq:Ermakov_eq_nu}
\end{align}
Because $W_t=W_t^{(\mu)}=W_t^{(\nu)}=1$, we obtain Eq.~(\ref{eq:Ermakov_eq}) from Eqs.~(\ref{eq:Ermakov_eq_mu}) and (\ref{eq:Ermakov_eq_nu}).


\section{Derivation of Ermakov equation~(\ref{eq:ermakov_pa}) by phase-amplitude method}
\label{sec:deriv_ermakoveq_phamp}

According to the phase-amplitude method~\cite{Andersson}, 
we rewrite $\mu_t$ and $\nu_t$ by using $\rho_t$ defined in Eq.~(\ref{eq:def_rho}), from which a phase function can naturally be defined.
Since $\dot\rho_t\mu_t-\rho_t\dot\mu_t=-\rho_t^2\diff{}{t}\frac{\mu_t}{\rho_t}$, the Wronskian given by Eq.~(\ref{eq:wronskian_mu}) can be rewritten as
\begin{align}
W^{(\mu)}_t
=
\frac{\rho_t^2}{\sqrt{\varOmega_{t_0}^{-1}-\bigl(\frac{\mu_t}{\rho_t}\bigr)^2}}\diff{}{t}\frac{\mu_t}{\rho_t}.
\end{align}
This can easily be integrated to obtain
\begin{align}
\mu_t
=
\frac{\rho_t}{\sqrt{\varOmega_{t_0}}}\sin\int_{t_0}^t{\rm d}t'\frac{W^{(\mu)}_{t'}}{\rho_{t'}^2}.
\label{eq:phase-amp_mu}
\end{align}
From Eqs.~(\ref{eq:wronskian}), (\ref{eq:def_rho}), and (\ref{eq:phase-amp_mu}), we also obtain
\begin{align}
\nu_t
=
\sqrt{\varOmega_{t_0}}\rho_t\cos\int_{t_0}^t{\rm d}t'\frac{W^{(\nu)}_{t'}}{\rho_{t'}^2}.
\label{eq:phase-amp_nu}
\end{align}
We now introduce the phase function $\theta_t$ defined as
\begin{align}
\theta_t
\equiv
\int_{t_0}^t{\rm d}t'\frac{W^{(\mu)}_{t'}}{\rho_{t'}^2}=\int_{t_0}^t{\rm d}t'\frac{W^{(\nu)}_{t'}}{\rho_{t'}^2}.
\label{eq:phase_def}
\end{align}
By differentiating Eq.~(\ref{eq:phase_def}) with respect to time $t$, we can represent the Wronskian with $\rho_t$ and $\theta_t$ as
\begin{align}
W_t=\rho_t^2\dot\theta_t.
\label{eq:wronskian_rho_theta}
\end{align}
By differentiating $\rho_t$ in Eq.~(\ref{eq:wronskian_rho_theta}) with respect to time $t$ twice, we have
\begin{align}
\ddot\rho_t+\biggl(-\frac{3}{4}\frac{\ddot\theta_t^2}{\dot\theta_t^2}+\frac{1}{2}\frac{\dddot\theta_t}{\dot\theta_t}\biggr)\rho_t
=
0.
\end{align}
Adding $\dot\theta_t^2\rho_t$ to both sides of the above equation and defining
\begin{align}
f_t
\equiv
\sqrt{\dot\theta_t^2-\frac{3}{4}\frac{\ddot\theta_t^2}{\dot\theta_t^2}+\frac{1}{2}\frac{\dddot\theta_t}{\dot\theta_t}},
\end{align}
we obtain the Ermakov equation:
\begin{align}
\ddot\rho_t+f_t^2\rho_t
=
\dot\theta_t^2\rho_t
=
\biggl(\frac{W_t}{\rho_t^2}\biggr)^2\rho_t
=
\frac{W_t^2}{\rho_t^3},
\end{align}
where we used Eq.~(\ref{eq:wronskian_rho_theta}).


\section{Derivation of Wronskian in Eqs.~(\ref{eq:wronskian_mu_omega}) and (\ref{eq:wronskian_nu_omega})}
\label{sec:deriv_wr}

By using Eqs.~(\ref{eq:rho_solution}), (\ref{eq:ELinv_mu}), and (\ref{eq:EL_inv_wron}), we obtain Eq.~(\ref{eq:wronskian_mu_omega}) as follows~\cite{Guasti09}:
\begin{align}
W^{(\mu)}_t
&=
\frac{2I^{(\mu)}_t}{W^{(\mu)}_t}
\notag\\
&=
\frac{\omega_{t_0}}{W^{(\mu)}_t}\biggl\{(\dot\rho_t\mu_t-\rho_t\dot\mu_t)^2+W_t^{(\mu)2}\frac{\mu_t^2}{\rho_t^2}\biggr\}
\notag\\
&=
\omega_{t_0}\biggl\{\biggl(\dot\mu_t-\frac{\dot\rho_t}{\rho_t}\mu_t\biggr)^2\frac{\rho_t^2}{W^{(\mu)}_t}+\mu_t^2\frac{W_t^{(\mu)}}{\rho_t^2}\biggr\}
\notag\\
&=
\omega_{t_0}\biggl\{\biggl(\dot\mu_t+\frac{\mu_t}{2}\frac{\dot\omega_t}{\omega_t}\biggr)^2\frac{1}{\omega_t}+\mu_t^2\omega_t\biggr\}
\notag\\
&=
\frac{2\omega_{t_0}}{\omega_t}\biggl\{\mathcal E^{(\mu)}_t+\biggl(\dot\mu_t+\frac{\mu_t}{2}\frac{\dot\omega_t}{\omega_t}\biggr)\frac{\mu_t}{2}\frac{\dot\omega_t}{\omega_t}\biggr\}.
\end{align}
Similarly, by using Eqs.~(\ref{eq:rho_solution}), (\ref{eq:ELinv_nu}), and (\ref{eq:EL_inv_wron}), we obtain Eq.~(\ref{eq:wronskian_nu_omega}) as follows:
\begin{align}
W^{(\nu)}_t
&=
\frac{2I^{(\nu)}_t}{W^{(\nu)}_t}
\notag\\
&=
\frac{1}{\omega_{t_0}W^{(\nu)}_t}\biggl\{(\dot\rho_t\nu_t-\rho_t\dot\nu_t)^2+W_t^{(\nu)2}\frac{\nu_t^2}{\rho_t^2}\biggr\}
\notag\\
&=
\frac{1}{\omega_{t_0}}\biggl\{\biggl(\dot\nu_t-\frac{\dot\rho_t}{\rho_t}\nu_t\biggr)^2\frac{\rho_t^2}{W^{(\nu)}_t}+\nu_t^2\frac{W_t^{(\nu)}}{\rho_t^2}\biggr\}
\notag\\
&=
\frac{1}{\omega_{t_0}}\biggl\{\biggl(\dot\nu_t+\frac{\nu_t}{2}\frac{\dot\omega_t}{\omega_t}\biggr)^2\frac{1}{\omega_t}+\nu_t^2\omega_t\biggr\}
\notag\\
&=
\frac{2}{\omega_t\omega_{t_0}}\biggl\{\mathcal E^{(\nu)}_t+\biggl(\dot\nu_t+\frac{\nu_t}{2}\frac{\dot\omega_t}{\omega_t}\biggr)\frac{\nu_t}{2}\frac{\dot\omega_t}{\omega_t}\biggr\}.
\end{align}


\section{Exact explicit form of transition probabilities in Eq.~(\ref{eq:trans_prob})}
\label{sec:exact_pmn}

Here, we derive an exact explicit form of the transition probabilities $P^{m,n}_{t,t_0}$ in Eq.~(\ref{eq:trans_prob}) as a function of time $t$ through the parameter $Q^{\rm TT}_t$ according to~\cite{Husimi,Deffner10}.
Since the probability generating function $\mathcal{P}^{u,v}_{t,t_0}$ cannot be expanded in powers of $u$ and $v$ in an explicit series, we introduce the following transition amplitude $U^{m,n}_{t,t_0}$~\cite{Husimi}:
\begin{align}
U^{m,n}_{t,t_0}
\equiv
\iint_{\mathbb{R}^2}{\rm d}x{\rm d}x_0\bket{m;\varOmega_t}{x}U_{t,t_0}(x|x_0)\bket{x_0}{n;\varOmega_{t_0}},
\label{eq:trans_amp}
\end{align}
where $P^{m,n}_{t,t_0}=|U^{m,n}_{t,t_0}|^2$.
By using the generating function of the $n$-th-degree Hermite polynomial
\begin{align}
e^{2xz-z^2}=\sum_{n=0}^\infty\frac{{\rm H}_n(x)}{n!}z^n,
\label{eq:genef_her}
\end{align}
\begin{widetext}
\noindent we have
\begin{align}
\sum_{n=0}^\infty\sqrt{\frac{2^n}{n!}}z^n\bket{x}{n;\varOmega_t}
=
\biggl(\frac{M\varOmega_t}{\pi\hbar}\biggr)^{1/4}\exp\biggl(-\frac{\zeta_tM\varOmega_t}{2\hbar}x^2\biggr)\exp\biggl(2\sqrt{\frac{M\varOmega_t}{\hbar}}xz-z^2\biggr).
\label{eq:use_gene_Hn}
\end{align}
We calculate the generating function of the transition amplitude $U^{m,n}_{t,t_0}$ as follows:
\begin{align}
\mathcal U^{u,v}_{t,t_0}
&\equiv
\sum_{m,n=0}^\infty\sqrt{\frac{2^{m+n-1}}{m!n!}}u^n v^m U^{m,n}_{t,t_0}
\notag\\
&=
\frac{1}{\sqrt{2}}\iint_{\mathbb R^2}{\rm d}x{\rm d}x_0 U_{t,t_0}(x|x_0)\sum_{m=0}^\infty\sqrt{\frac{2^m}{m!}}v^m\bket{m;\varOmega_t}{x}\sum_{n=0}^\infty\sqrt{\frac{2^n}{n!}}u^n\bket{x_0}{n;\varOmega_{t_0}}
\notag\\
&=
\frac{M}{2\pi\hbar}\frac{(\varOmega_t\varOmega_{t_0})^{1/4}}{\sqrt{{\rm i}\mu_t}}e^{-(u^2+v^2)}\int_{\mathbb R^2}{\rm d}\vec x \exp\biggl[-\frac{M}{2\hbar}
(\underbrace{%
\vec x\cdot B\vec x-2\vec b\cdot\vec x
\vphantom{\frac{M}{2\hbar}}
}%
_{\hspace{-5em}(\vec x-B^{-1}\vec b)\cdot B(\vec x-B^{-1}\vec b)-\vec b\cdot B^{-1}\vec b\hspace{-5em}
}%
)\biggr]
\notag\\
&=
\frac{(\varOmega_t\varOmega_{t_0})^{1/4}}{\sqrt{{\rm i}\mu_t\det B}}e^{-(u^2+v^2)}\exp\biggl(\frac{M}{2\hbar}\vec b\cdot B^{-1}\vec b\biggr)
\notag\\
&=
\frac{(\varOmega_t\varOmega_{t_0})^{1/4}}{\sqrt{{\rm i}\chi_t^{(-)}}}\exp\Biggl(\frac{\chi_t^{(+)} u^2-4{\rm i}\sqrt{\varOmega_t\varOmega_{t_0}}uv+\chi^{(+)*}_t v^2}{\chi_t^{(-)}}\Biggr).
\label{eq:Uuv}
\end{align}
In the above, we have defined the following quantities:
\begin{align}
&\vec x
\equiv
\begin{pmatrix}
x\vphantom{\frac{\dot\mu_t}{\mu_t}}\\
x_0\vphantom{\frac{\nu_t}{\mu_t}}
\end{pmatrix},
\quad
B
\equiv
\begin{pmatrix}
\varOmega_t-{\rm i}\frac{\dot\mu_t}{\mu_t}&\frac{{\rm i}}{\mu_t}\\
\frac{{\rm i}}{\mu_t}&\varOmega_{t_0}-{\rm i}\frac{\nu_t}{\mu_t}
\end{pmatrix},
\\
&\vec b
\equiv
2\sqrt{\frac{\hbar}{M}}
\begin{pmatrix}
\sqrt{\varOmega_t}v\vphantom{\frac{\dot\mu_t}{\mu_t}}\\
\sqrt{\varOmega_{t_0}}u\vphantom{\frac{\nu_t}{\mu_t}}
\end{pmatrix},
\\
&\chi_t^{(\pm)}
\equiv
\varOmega_{t_0}(\varOmega_t \mu_t-{\rm i}\dot\mu_t)\pm {\rm i}(\varOmega_t \nu_t-{\rm i}\dot\nu_t).
\end{align}
Note that
$\det B=\frac{\chi_t^{(-)}}{\mu_t}$
and
$|\chi_t^{(\pm)}|^2=2\varOmega_t\varOmega_{t_0}(\mathcal Q_t\mp1)$,
where
\begin{align}
\mathcal Q_t
\equiv
\varOmega_{t_0}\frac{\mathcal E^{(\mu)}_t}{\varOmega_t}+\varOmega_{t_0}^{-1}\frac{\mathcal E^{(\nu)}_t}{\varOmega_t},
\end{align}
which agrees with $Q^{\rm TT}_t$ in Eq.~(\ref{eq:Q_TT_E}) if $\dot\omega_{t_0}=0$ is imposed.
For a usual QPO in the absence of $\hat H^{\rm cd}_t$, $\mathcal Q_t$ is identified as Husimi's measure of adiabaticity $Q^*_t$.
From the symmetric property of $\mathcal P^{-u,-v}_{t,t_0}=\mathcal P^{u,v}_{t,t_0}$, 
we find $P^{m,n}_{t,t_0}=|U^{m,n}_{t,t_0}|^2=0$ if $m$ and $n$ are of different parity.
Then, by expanding Eq.~(\ref{eq:Uuv}) explicitly in powers of $u$ and $v$, we can obtain the matrix elements of $U^{m,n}_{t,t_0}$ as
\begin{align}
U^{m,n}_{t,t_0}
=
(2\varOmega_t\varOmega_{t_0})^{1/4}\sqrt{\frac{m!n!}{2^{m+n-1}}}\sqrt{\frac{\chi_t^{(+)*m}\chi_t^{(+)n}}{{\rm i}\chi_t^{(-)m+n+1}}}\sum_{s\ge 0}^{{\min(m,n)}}\frac{2^s\bigl(\frac{2}{1-\mathcal Q_t}\bigr)^{s/2}}{s!\bigl(\frac{m-s}{2}\bigr)!\bigl(\frac{n-s}{2}\bigr)!}.
\label{eq:Upower}
\end{align}
By applying the selection rule $m-n\in 2\mathbb{Z}$, the number $s$ satisfies $s\in 2\mathbb{N}_0$ for $m,n\in 2\mathbb{N}_0$, and $s\in 2\mathbb{N}_0+1$ for $m,n\in 2\mathbb{N}_0+1$, where $\mathbb{N}_0\equiv\mathbb{N}\cup\{0\}$.
The explicit expression for the matrix elements of $U^{m,n}_{t,t_0}$ reads for even elements and odd elements~\cite{Deffner10}, respectively, as
\begin{align}
U^{2k,2l}_{t,t_0}
&=
\frac{(2\varOmega_t\varOmega_{t_0})^{1/4}}{k!l!}\sqrt{\frac{(2k)!(2l)!}{2^{2(k+l)-1}}}\frac{\chi_t^{(+)*k}\chi_t^{(+)l}}{\chi_t^{(-)k+l}\sqrt{{\rm i}\chi_t^{(-)}}}{}_2F_1\biggl(-k,-l;\frac{1}{2};\frac{2}{1-\mathcal Q_t}\biggr),
\\
U^{2k+1,2l+1}_{t,t_0}
&=
-\frac{(2\varOmega_t\varOmega_{t_0})^{1/4}}{k!l!}\sqrt{\frac{(2k+1)!(2l+1)!}{2^{2(k+l)+1}
}}\frac{\chi_t^{(+)*k}\chi_t^{(+)l}}{\chi_t^{(-)k+l}\sqrt{{\rm i}\chi_t^{(-)}}}\frac{|\chi_t^{(+)}|}{\chi_t^{(-)}}\sqrt{\frac{2}{1-\mathcal Q_t}}{}_2F_1\Bigl(-k,-l;\frac{3}{2};\frac{2}{1-\mathcal Q_t}\Bigr),
\end{align}
where $k,l\in\mathbb N_0$ and ${}_2F_1$ is Gauss's hypergeometric function.
We finally obtain the explicit closed form of transition probabilities as functions of the parameter $\mathcal Q_t$, which reads for even elements and odd elements, respectively, as
\begin{align}
P^{2k,2l}_{t,t_0}
&=
\frac{(2k-1)!!(2l-1)!!}{(2k)!!(2l)!!}\sqrt{\frac{2}{\mathcal Q_t+1}}\biggl(\frac{\mathcal Q_t-1}{\mathcal Q_t+1}\biggr)^{k+l}{{}_2F_1}^2\biggl(-k,-l;\frac{1}{2};\frac{2}{1-\mathcal Q_t}\biggr),
\\
P^{2k+1,2l+1}_{t,t_0}
&=
\frac{(2k+1)!!(2l+1)!!}{(2k)!!(2l)!!}\biggl(\frac{2}{\mathcal Q_t+1}\biggr)^{3/2}\biggl(\frac{\mathcal Q_t-1}{\mathcal Q_t+1}\biggr)^{k+l}{{}_2F_1}^2\biggl(-k,-l;\frac{3}{2};\frac{2}{1-\mathcal Q_t}\biggr).
\end{align}
\end{widetext}


\end{document}